# Unveiling coupling between carrier recombination pathways in semiconductors via mechanical damping

Mingyu Xie[1], Ruitian Chen[1], Jiaze Wu[1], Kaiqi Qiu[1], Mingqiang Li[1], Huicong Chen[1], Kai Huang[1], Yu Zou[1,a)]

[1]Department of Materials Science and Engineering, University of Toronto, Toronto, ON M5S 3E4, Canada

## Abstract

The total rate of carrier recombination in semiconductors has conventionally been expressed using an additive model, $r_{total} = \sum r_i$, which rules out the interactions between recombination pathways. Here we challenge this paradigm by demonstrating pathway coupling using a new method, light-induced mechanical absorption spectroscopy (LIMAS), which probes recombination dynamics via mechanical damping. We show that the total recombination rate in zinc sulfide (ZnS), a model semiconductor material, follows a multiplicative weighting model, $r_{total} \propto \prod {r_i}^{w_i}$ with $\sum w_i = 1$. Under steady-state and switch-on illuminations, the weighting factors $w_i$ for each pathway—direct, trap-assisted, and sublinear—are dictated by the carrier generation mechanisms: Interband transition favors direct recombination; Single defect level-mediated generation promotes trap-assisted recombination; Generation involving multiple saturated defect levels gives rise to sublinear recombination. Taking the additive and multiplicative models as two limiting cases, we propose a generalized power-mean recombination model, $r_{total} \propto \left(\sum w_i r_i^p\right)^{1/p}$ with $p \in (0,1]$, where $p$ quantifies the degree of coupling between recombination pathways.

[a)] Author to whom all correspondence should be addressed; Email: mse.zou@utoronto.ca (Y.Z.)

## 1. Introduction

Semiconductors are widely used in LEDs [1], photodetectors [2], and solar cells [3], with their optoelectronic properties primarily governed by carrier dynamics, particularly recombination processes [4-7]. Multiple recombination pathways, such as direct, trap-assisted [8, 9], sublinear [10], excitonic [11], and Auger recombination [12], have been proposed to describe the recombination mechanisms in semiconductors. These coexisting pathways have been characterized using techniques including *I-V* testing [13], transient absorption [14, 15], and power-dependent/time-resolved photoluminescence [16] for a wide range of materials, from conventional semiconductors [12, 17] to hybrid perovskites [18, 19]. In most of these studies, the recombination dynamics are analyzed using a simple additive model, $r_{total} = \sum r_i$, where $r_{total}$ is the total recombination rate and $r_i$ is the rate of each pathway [20-22]. Although this additive model, also known as the ABC model [20], has contributed to a fundamental understanding of semiconductors over the past few decades [6, 23], significant discrepancies between its predictions and experimental results have been reported under various conditions [24-27], implying that additional mechanisms beyond the additive picture are involved. Recent studies on nanocrystals suggest that recombination pathways can couple rather than act independently. For example, in $FAPbBr_3$ nanocrystals, excitonic and hot carrier recombination dominate at different wavelengths [28], whereas in InP/CdS core-shell quantum dots, the balance between Auger and surface recombination depends on the shell thickness [29]. For bulk semiconductors, however, direct evidence for such pathway coupling remains scarce.

Here we introduce light-induced mechanical absorption spectroscopy (LIMAS), a technique that probes recombination dynamics in piezoelectric semiconductors via mechanical damping. Unlike traditional optical or electrical techniques, LIMAS detects both radiative and non-radiative recombination while avoiding parasitic effects from electrodes [30, 31]. Our tests on zinc sulfide show that the total recombination rate under illumination follows a multiplicative model, $r_{total} \propto \prod {r_i}^{w_i}$ with $\sum w_i = 1$, which corresponds to the $p \to 0$ limit of the generalized power-mean model, $r_{total} \propto \left(\sum w_i r_i^p\right)^{1/p}$, and represents strong coupling between recombination pathways. In contrast, the previous additive ABC model ($p$=1 limit) depicts decoupled pathways.

## 2. Materials and Methods

### 2.1 Single Crystals

Two high-purity cubic ZnS single crystals, denoted ZnS (i) and ZnS (ii), were commercially obtained. The crystals were supplied as wafers (~10 mm × 10 mm × 1 mm) with optical-grade surfaces. The large faces were oriented parallel to the (110) crystallographic plane. The as-received ZnS (i) was annealed in vacuum (500 °C, 120 h), whereas ZnS (ii) was annealed in a sulfur atmosphere (500 °C, 120 h) [32].

### 2.2 LIMAS

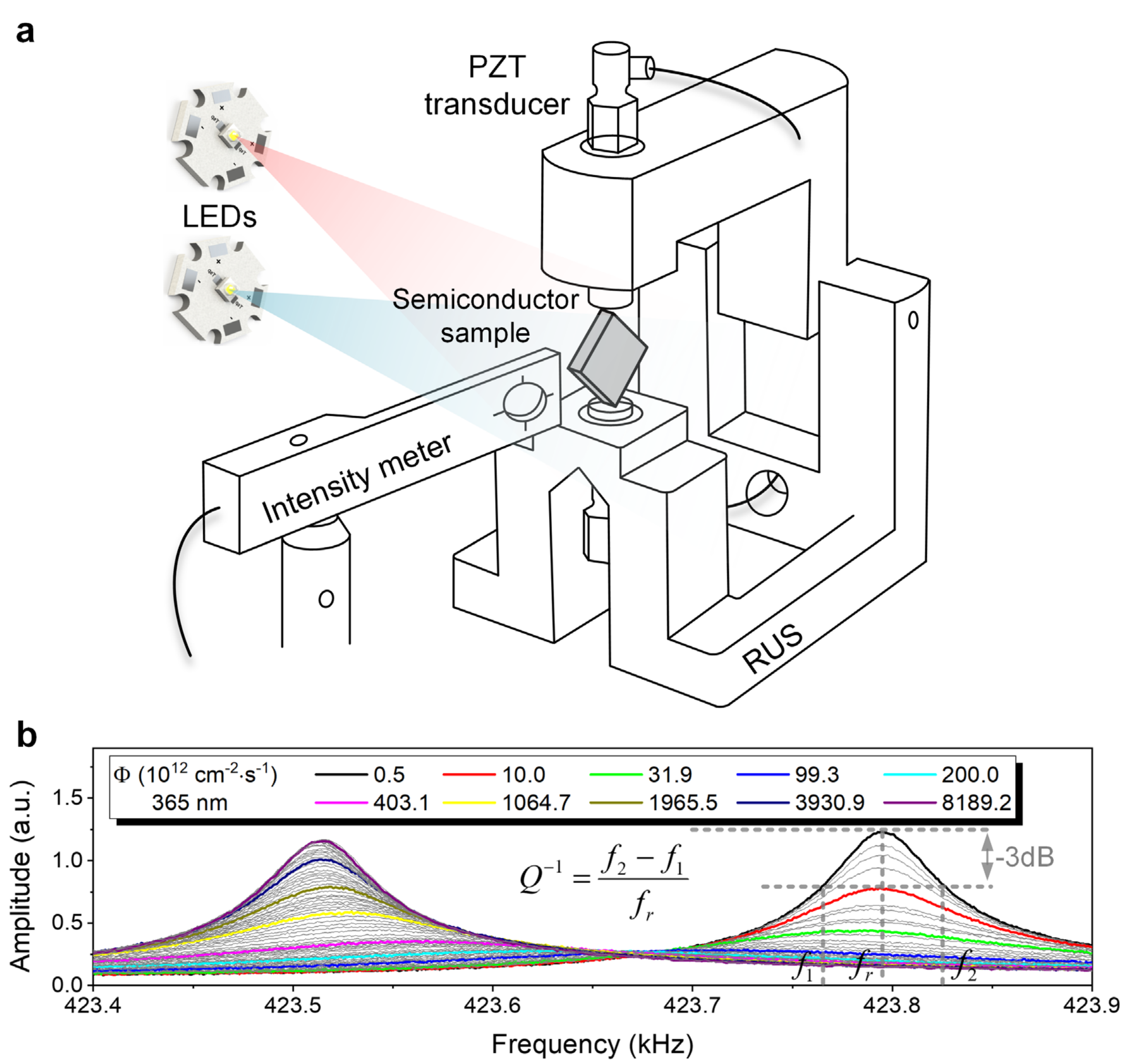

**Fig. 1: LIMAS setup and extraction of the mechanical response. a,** Schematic of the light-induced mechanical absorption spectroscopy (LIMAS). **b,** Representative amplitude-frequency curves of ZnS (i) measured under 365 nm illumination at different photon fluxes. The resonance

frequency $f_r$ is determined from the peak position, while $f_1$ and $f_2$ denote the lower and upper -3 dB frequencies used to determine the mechanical damping $Q^{-1}$.

The LIMAS setup used in this work is shown in Fig. 1**a**. The system utilizes resonant ultrasound spectroscopy [33, 34] (RUS, Alamo Creek Engineering) to measure damping ($Q^{-1}$) and resonance frequency changes ($\Delta f_r$). The semiconductor sample was lightly clamped at two opposite corners by a pair of PZT transducers, which were mounted on a vibration isolation stage to minimize external disturbances. The transmitting PZT transducer excited elastic vibrations over a frequency range from kHz to MHz. The receiving PZT transducer recorded the amplitude-frequency response, in which a resonance peak appeared when the excitation frequency matched one of the sample's eigenfrequencies. As illustrated in Fig. 1**b**, $f_r$ was determined from the peak position, while the damping was calculated from the -3 dB bandwidth using $Q^{-1} = (f_2 - f_1)/f_r$. For illumination, various LEDs with emission wavelengths ranging from 300 nm to 1100 nm were used. The photon flux was measured using a Si-based photodiode power sensor (S120VC, 200 -1100 nm, Thorlabs) with a resolution of 1 nW. During the test, the photodiode power sensor was positioned close to the sample along the optical axis to estimate the incident photon flux at the illuminated region. Experiments were conducted under both steady-state illumination (monochromatic and bichromatic) and transient conditions, with a time resolution of 1 s for the transient measurements.

### 2.3 Photoluminescence spectra

Photoluminescence contour maps were measured using a Horiba spectrofluorometer (FluoroMax Plus) equipped with a xenon lamp as the excitation source. Measurements were conducted at room temperature with a 2 nm emission step size, 5 nm excitation step size, 0.1 s integration time, and (2×2) entrance and exit slits. For time-dependent light emission measurements, a 365 nm LED was used as the excitation source, and the spectrofluorometer recorded the emission during the switch-on and switch-off of the LED.

### 2.4 Transmission spectra

A Lambda 365 Spectrometer was used to measure optical transmission between 200 nm and 1000 nm. The measurements were carried out at room temperature, with a 5 nm interval, 0.1 s integration time, and 2.0 nm bandwidth.

### 2.5 Photon flux-dependent photoluminescence quantum yield (PLQY)

PLQY measurements were performed on ZnS (i) at room temperature using LEDs, an integrating sphere, and a Si-based photodiode power sensor (S120VC, Thorlabs). The sample was excited at 340 and 365 nm under varying photon fluxes. For each excitation condition, the emitted and absorbed photon fluxes were measured, and the absolute PLQY was calculated as the ratio of emitted photons to absorbed excitation photons.

## 3. Results and analysis

### 3.1 Theoretical framework of LIMAS

Fig. 2 illustrates the theoretical framework of LIMAS. In a uniformly illuminated yet mechanically static homogeneous compound semiconductor, conduction band electrons ($n = n_0 + \Delta n$), valence band holes ($p = p_0 + \Delta p$), and localized states electrons remain uniformly distributed throughout the sample (Fig. 2**b**). Here, $n_0$ and $p_0$ are thermal equilibrium carrier concentrations, $\Delta n$ and $\Delta p$ denote photo-generated excess carriers, and $A$, $B$, and $C$ denote different regions of the sample (Fig. 2**a**). Under this condition, no drift or diffusion current occurs. However, when the sample vibrates at one of its piezoelectrically active resonant eigenmodes (Supplementary Note 3 [35]), piezoelectricity induces a weak internal alternating potential ($10^{-9}$-$10^{-7}$ V), perturbing the electron and hole concentrations ($\Delta n'$ and $\Delta p'$, Fig. 2**c**). This perturbation generates an internal alternating current, leading to mechanical damping ($Q^{-1}$) and resonance frequency change ($\Delta f_r/f_r$) [36, 37], which can be expressed as follows (see Supplementary Note 4 for detailed derivation [35]):

$$Q^{-1} = \frac{e^2}{c\varepsilon}\frac{\omega_c/\omega}{1+(\omega_c/\omega)^2},\quad \frac{\Delta f_r}{f_r} = -\frac{e^2}{2c\varepsilon}\frac{(\omega_c/\omega)^2}{1+(\omega_c/\omega)^2} \tag{1}$$

where $c$, $e$, and $\varepsilon$ are the elastic, piezoelectric and dielectric constants, respectively. $\omega = 2\pi f_r$ represents the mechanical vibration frequency and can be considered constant during tests since $\Delta f_r/f_r \ll 1$. $\omega_c = q(\mu_n \Delta n + \mu_p \Delta p)/(f\varepsilon)$ is the conductivity frequency, with $\mu_n$, $\mu_p$ as carrier mobilities, $q$ as the elementary charge, and $f$ as the fraction of net space charge contributed by mobile carriers (generally $f \leq 1$, here we set $f=1$ by assuming mid-gap states remain unchanged). Therefore, illumination affects the mechanical response mainly by changing the carrier concentration and hence the conductivity frequency $\omega_c$. According to Eq. (1), $Q^{-1}$ reaches a maximum when $\omega = \omega_c$, whereas $\Delta f_r/f_r$ evolves monotonically from the unrelaxed to the relaxed state.

For semiconductors under steady-state illumination with a single given recombination pathway, the recombination rate $r = \alpha\beta\Phi \propto (\Delta n)^{\gamma}$ (Fig. 2**d**), where $\Phi$ is the photon flux, and $\alpha$, $\beta$ denote the absorption coefficient and carrier generation quantum yield, respectively. The recombination index $\gamma$ takes values of 3, 2, 1, and <1 for Auger, direct, trap-assisted, and sublinear recombination, respectively (Supplementary Note 5 [35]). Trap-assisted recombination, also known as SRH recombination [8, 9], occurs via a single unsaturated defect level (i.e., recombination center, RC, in Fig. 2**d**), whereas sublinear recombination involves multiple saturated defect levels [10, 38]. Fig. 2**e** shows the calculated steady-state photon flux-damping ($\Phi$-$Q^{-1}$) and photon flux-resonance frequency change curves ($\Phi$-$\Delta f_r/f_r$) for each recombination pathway, based on Eq.(1) (see Supplementary Note 6 for calculation parameters [35]). All curves exhibit standard anelastic behavior: the $\Phi$-$Q^{-1}$ curves follow a Lorentzian peak shape, with the characteristic width ($\gamma$) representing the detailed recombination pathway, and the peak position ($\Phi_P$) indicating the recombination rate. A peak shift to higher (lower) photon flux region suggests an increased (decreased) recombination rate. Meanwhile, the $\Phi$-$\Delta f_r/f_r$ curves display a transition from the unrelaxed to the relaxed state. Under transient illumination with a single given recombination pathway, the time-dependent $Q^{-1}$ can also be calculated by first solving the rate equation: $\frac{d\Delta n}{dt} = -r + \alpha\beta\Phi$, and then substituting the resulting $\Delta n$ into Eq. (1).

In practical semiconductors, multiple pathways coexist. If these pathways operate independently, the total recombination rate is conventionally written as the additive model [20], $r_{total} = \sum r_i$, which corresponds to the $p$=1 limit of the generalized power-mean model (Supplementary Note 14 [35]). Because different pathways generally have different dependences on carrier concentration, i.e., $r_i \propto \Delta n^{\gamma_i}$, their summation does not, in general, preserve a single power-law dependence on Δn. Therefore, the additive ABC model is expected to lead to deviations from standard anelastic behavior in both $\Phi$-$Q^{-1}$ and $\Phi$-$\Delta f_r/f_r$ curves (Supplementary Note 8 [35]).

However, if a standard anelastic response is maintained, evidenced by $\Phi$-$Q^{-1}$ being well described by a Lorentzian peak, it indicates that $r_{total}$ can be effectively described by a single power-law relationship with Δn. Such behavior is naturally obtained from the multiplicative weighting model, i.e., $p\to0$ limit of the power-mean model:

$$\begin{cases} r_{\text{total}} \propto \prod r_i^{w_i}, \quad \sum w_i = 1, \quad w_i \in [0,1] \\ r_i \propto \Delta n^{\gamma_i}, \ \gamma_i = 3, 2, 1, \text{and} < 1 \text{ for (iii), (i), (ii), and (iv)} \end{cases} \tag{2}$$

Combining these relations gives $r_{total} \propto \Delta n^{\gamma_{eff}}$, where $\gamma_{eff} = \sum w_i \gamma_i$. Thus, in this condition, the experimentally extracted recombination index $\gamma_{eff}$ reflects the weighted contribution of the coexisting recombination pathways. Here, $w_i$ is the weighting factor for each pathway (e.g., Auger (iii), direct (i), trap-assisted (ii), sublinear (iv), etc.) and varies with both the sample's internal defects and external conditions such as photon energy.

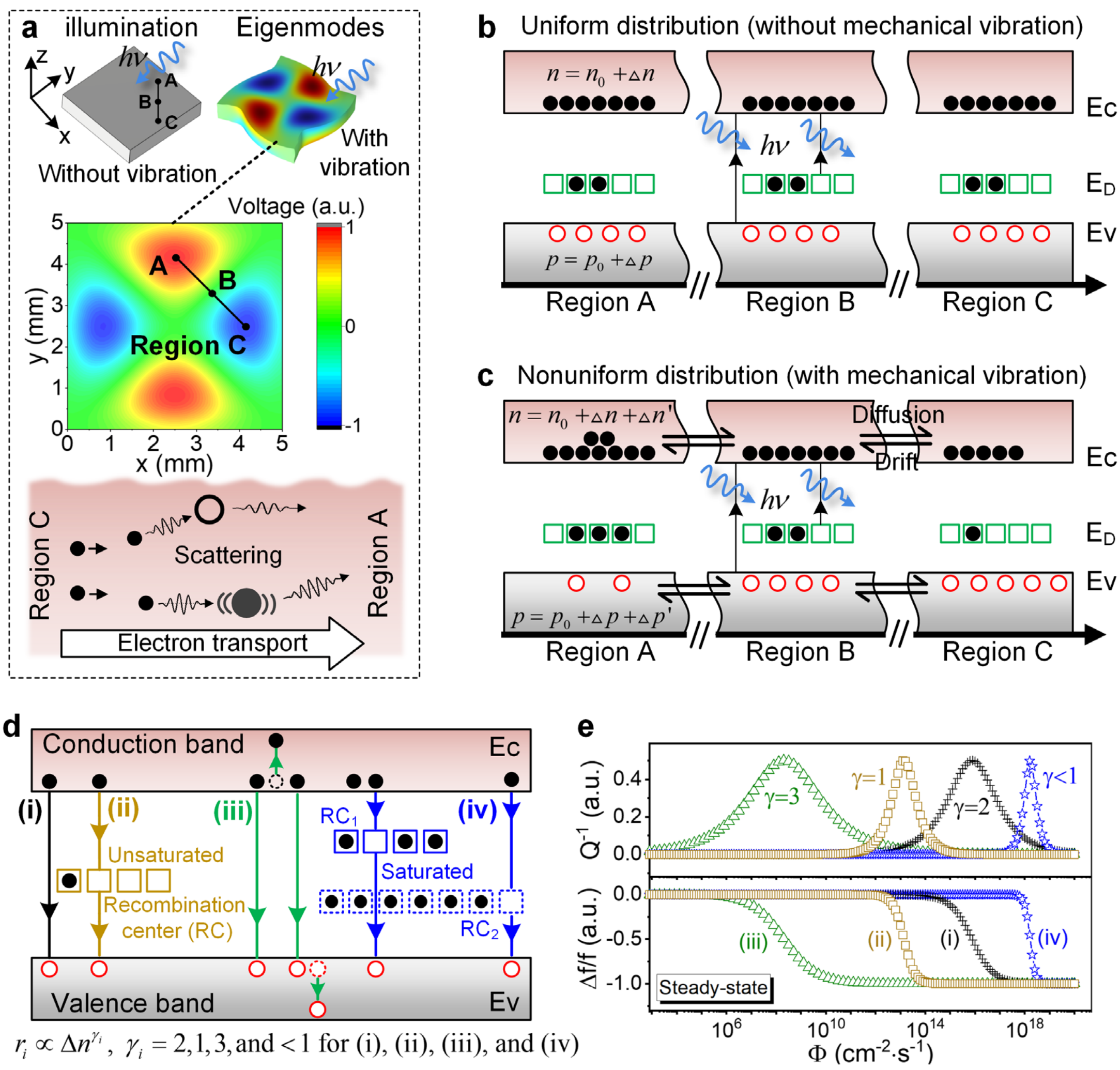

**Fig. 2: Theoretical framework of LIMAS for probing recombination dynamics in piezoelectric semiconductors. a,** Alternating potentials and carrier transport in a uniformly illuminated semiconductor under piezoelectrically active resonant eigenmodes. **b and c,** Distributions of conduction band electrons (black dots), valence band holes (red hollow dots), and localized states electrons across different regions (A, B, and C) of the sample without and with mechanical vibration. **d,** Power-law relationship between recombination rate ($r$) and carrier concentration (Δn) under steady-state illumination for each recombination pathway: (i) Direct, $r \propto \Delta n^2$; (ii) Trap-assisted, $r \propto \Delta n$; (iii) Auger, $r \propto \Delta n^3$; (iv) Sublinear, $r \propto \Delta n^\gamma$, $\gamma \in (0,1)$. **e,** Calculated steady-state photon flux-damping ($\Phi$-$Q^{-1}$) and photon flux-resonance frequency change curves ($\Phi$-$\Delta f_r/f_r$) for each single pathway, where $\gamma$ is the exponent of the carrier density in the recombination rate $r$ and determines the characteristic width of the $\Phi$-$Q^{-1}$ curve.

### 3.2 The coupling of recombination pathways under steady-state illumination

To demonstrate the coupling between recombination pathways, we performed LIMAS measurements on vacuum annealed ZnS (i) and sulfur annealed ZnS (ii). These crystals were exposed to steady-state monochromatic light (wavelength ranging from 320 to 500 nm) with photon flux $\Phi = 10^{11}$-$10^{18}\ cm^{-2} \cdot s^{-1}$, yielding carrier densities of $10^{7}$-$10^{13}\ cm^{-3}$ (Supplementary Note 8 [35]). The mechanical damping ($Q^{-1}$) and resonance frequency change ($\Delta f_r$) of the samples were measured using resonant ultrasound spectroscopy [33] at a resonant frequency ($f_r$) of approximately 420 kHz (Supplementary Notes 2 and 7 [35]). Fig. 3**a, b, d, e** show that the measured $\Phi$-$Q^{-1}$and $\Phi$-$\Delta f_r$ curves exhibit well-defined anelastic responses, as indicated by the Lorentzian damping peaks across various wavelengths. The insets above Fig. 3**a** and **b** compare representative experimental data under 365 nm illumination with fits obtained using the multiplicative model and the additive model. The multiplicative weighting model reproduces both the damping and resonance frequency responses more accurately than the additive model, providing strong evidence for pathway coupling. Notably, this behavior is not universal. GaN:C exhibits asymmetric damping peaks that are better described by the additive model, whereas the response of GaAs:Cr lies between the $p \to 0$ and $p$=1 limits, demonstrating that LIMAS can distinguish different degrees of pathway coupling (Supplementary Note 9 [35]).

All $\Phi$-$Q^{-1}$ curves of ZnS in Fig. 3**a** and **d** were subsequently fitted using the multiplicative weighting model (see Supplementary Note 8 for fitting details [35]), yielding the photon flux at the damping peak ($\Phi_P$) and the corresponding $\gamma_{eff}$ values as functions of wavelength, as shown in Fig. 3**c** and **f**. When the photon energy exceeds the bandgap energy ($E_g$ = 3.6 eV, ~345 nm, Fig. 3**h**), intrinsic absorption and interband transitions dominate. In this regime, the $\gamma_{eff}$ values for both ZnS (i) and (ii) gradually approach 2, indicating an increasing weight for direct recombination, i.e., $w_{direct}$ increases. Auger recombination is ruled out here, as it primarily occurs in heavily doped or strongly pumped materials [39]. The inferred recombination mechanisms, based on the extracted $\gamma_{eff}$ values, are annotated in Fig. 3**c**, **f** and **g**, with larger font sizes representing greater weighting factors.

For extrinsic absorption, the recombination behaviors differ. For ZnS (i) in the 350-440 nm range, carrier generation is primarily mediated by a single defect level. Both $\gamma_{eff}$ and $\Phi_P$ remain stable at approximately 1.2 and $1.5 \times 10^{14}\ cm^{-2} \cdot s^{-1}$, suggesting unchanged $w_i$. According to the multiplicative model, trap-assisted recombination dominates in this regime with a weighting factor $w_{trap}$ ≈0.8, while direct recombination ($w_{direct}$) contributes ≈ 0.2. As monochromatic light shifts beyond 440 nm but remains below 520 nm, $\gamma_{eff}$ drops below 1, and the damping peak shifts rapidly toward higher photon flux. This indicates the onset and progressive enhancement of sublinear recombination, i.e., $w_{sub}$ increases. Photoluminescence mapping (Fig. 3**g**) of ZnS (i) further supports the variation of $w_i$ with photon energies, and assigns the observed green light emission (~520 nm) to trap-assisted recombination. Compared to ZnS (i), ZnS (ii) exhibits slightly larger $\Phi_P$ and $\gamma_{eff}$ values under extrinsic absorption (Fig. 3**f**). Fig. 3**g** and **h** also show that ZnS (ii) has weaker green emission and lower absorption coefficients, indicating a lower concentration of specific point defects due to the sulfurization. DFT calculations (Supplementary Note 13 [35]) identify this defect as the zinc antisite, which introduces a deep level within the bandgap, serving as the recombination center that facilitates trap-assisted and sublinear recombination. Consequently, in ZnS (ii), these defect-related pathways are weaker, while direct recombination remains a non-negligible competitor under extrinsic absorption, leading to observed larger $\Phi_P$ and $\gamma_{eff}$.

Photon flux-dependent PLQY measurements on ZnS (i) (Fig. 3**i**) further support the multiplicative weighting model. The PLQY remains nearly constant over a wide excitation range, suggesting that the radiative fraction of the total recombination process is approximately unchanged, as expected for the $p \to 0$ limit of the power-mean model (Supplementary Note 14 [35]). By combining the PLQY results with the pathway weights extracted from Fig. 3**c**, we estimate that trap-assisted recombination has a radiative fraction of approximately 5% under these conditions.

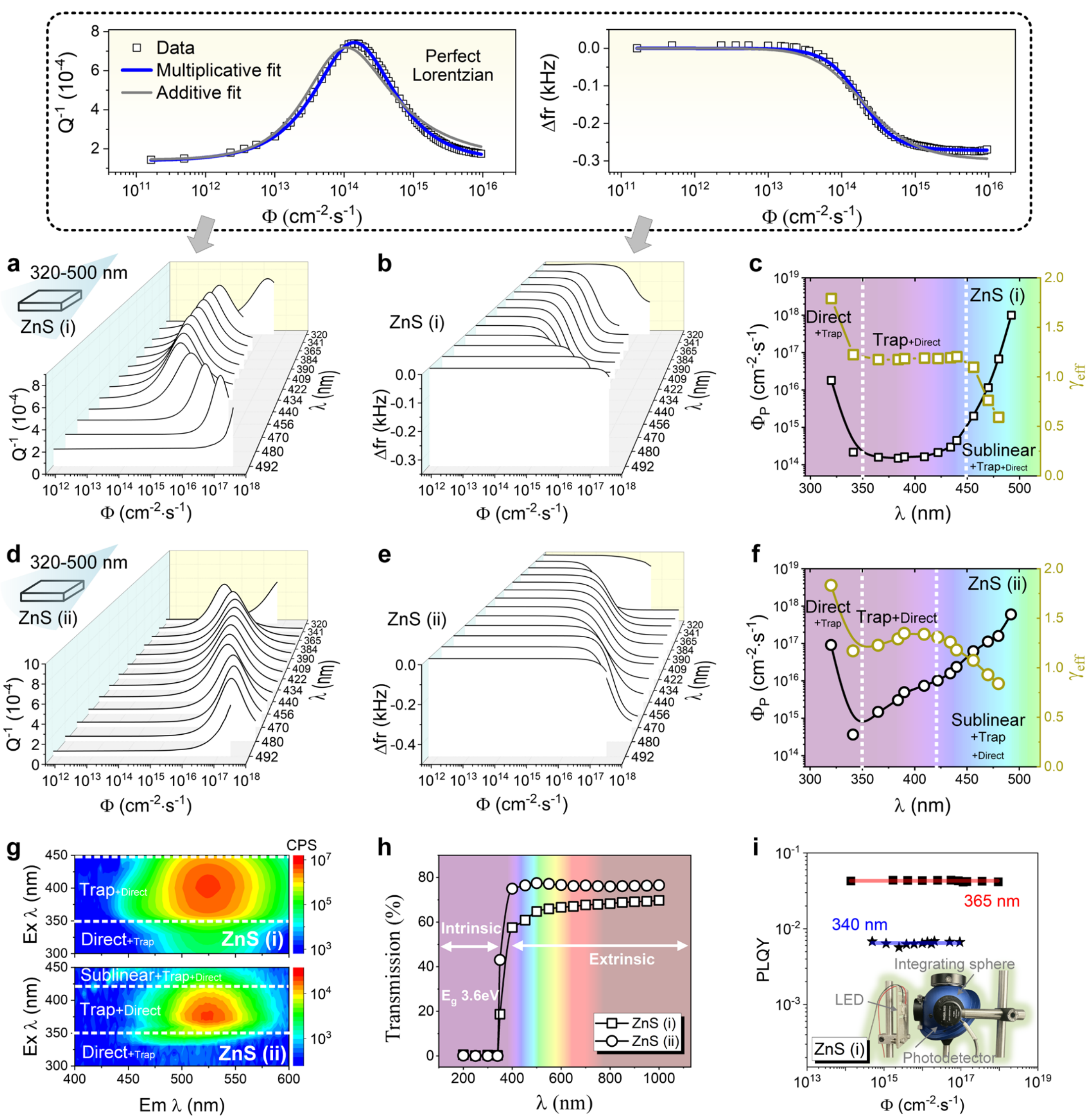

**Fig. 3: Responses of ZnS to steady-state monochromatic light**. **a, b, d, e,** Photon flux-damping ($\Phi$-$Q^{-1}$) and photon flux-resonance frequency change curves ($\Phi$-$\Delta f_r$) of ZnS (i) and (ii) under steady-state monochromatic light with varying wavelengths ($\lambda$). The insets above **a and b** compare representative experimental data under 365 nm illumination with fits obtained using the multiplicative model and the additive model. **c and f,** Photon flux at the damping peak ($\Phi_P$), characteristic width ($\gamma_{eff}$) of the damping peak, and the inferred recombination mechanisms based on $\gamma_{eff}$, for ZnS (i) and (ii) under monochromatic light with different wavelengths ($\lambda$). **g,** Photoluminescence contour maps of ZnS (i) and (ii), with excitation wavelengths ($E_x$ $\lambda$) ranging from 300 to 450 nm and detected emission wavelengths ($E_m$ $\lambda$) from 400 to 600 nm. **h,** Transmission spectra of ZnS (i) and (ii). **i,** Photon flux-dependent PLQY of ZnS (i) under 340 and 365 nm excitation; inset shows the integrating sphere configuration used for PLQY measurements.

We also introduce a second light source to examine its effect on the weighting factor $w_i$ established under monochromatic light. Here, a constant-intensity 365 nm UV light combined with an additional light source (400-1100 nm) with varying photon flux $\Phi_{additional}$ ($10^{12}\ cm^{-2}\cdot s^{-1}$ to $10^{17}\ cm^{-2}\cdot s^{-1}$) is applied (Fig. 4**a-d**). The UV light serves as the background because it mainly induces trap-assisted recombination (Fig. 3**c** and **f**), and the additional light mainly influences this pathway, as evidenced by the photoluminescence quenching contour map in Supplementary Note 10 [35]. The UV light establishes the baseline Δn, Δp, $Q^{-1}$, and $\Delta f_r$ (hollow dots in Fig. 4**a**), while the additional light induces two opposite effects: (1) Injection, which increases Δn and Δp, thereby reducing both $Q^{-1}$ and $\Delta f_r$ (blue arrows). (2) Quenching [38, 40], which decreases Δn and Δp, restoring $Q^{-1}$ and $\Delta f_r$ (red arrows). Fig. 4**b-d** present the $Q^{-1}$ and photoluminescence responses. In ZnS (i), the additional light with wavelengths below 440 nm induces only injection (Fig. 4**b**). Between 440 and 500 nm, injection and quenching compete, producing peaks in both the $Q^{-1}$ and $\Delta f_r$ curves (Supplementary Note 11 [35]). Beyond 520 nm, quenching dominates, recovering $Q^{-1}$, increasing $\Delta f_r$ (Supplementary Note 11 [35]), reducing green light emission [40], and generating a weak orange emission (Fig. 4**d**). ZnS (ii) follows similar trends, but quenching occurs at a shorter wavelength, about 430 nm (Fig. 4**c**).

By correlating the characteristic wavelengths at which quenching (Fig. 4**b** and **c**) and sublinear recombination (Fig. 3**c** and **f**) occur, we infer that quenching results from the onset of sublinear recombination. This process reduces $w_{trap}$ while increasing $w_{sub}$, thereby altering light emission and increasing $r_{total}$. To further validate this, damping contour maps (Fig. 4**e**) were recorded for ZnS (i). Results show that quenching, represented by the 'bending' in the contour maps, coincides with the peak narrowing ($\gamma_{eff}$ decreases), reflecting the onset and progressive enhancement of sublinear recombination. Moreover, the narrowed damping peaks retain perfect Lorentzian shapes (see waterfall plot in Fig. 4**f** and Supplementary Note 8 [35]), confirming that $r_{total}$ still adheres to the multiplicative weighting model, i.e., the $p \to 0$ limit of the power-mean model, under bichromatic illumination.

Analysis of $w_i$ under steady-state illumination shows that electron-hole pairs preferentially recombine through the pathways by which they were generated. For $E_{photon}>E_g$, interband transitions dominate, promoting direct recombination. Slightly below the bandgap (0.8-1$E_g$), carrier generation is primarily governed by a single defect level, favoring trap-assisted recombination. At much lower energies (0.66-0.8$E_g$) or with additional long wavelength illumination ($E_{photon}$<0.8$E_g$), saturation of multiple defect levels induces sublinear recombination.

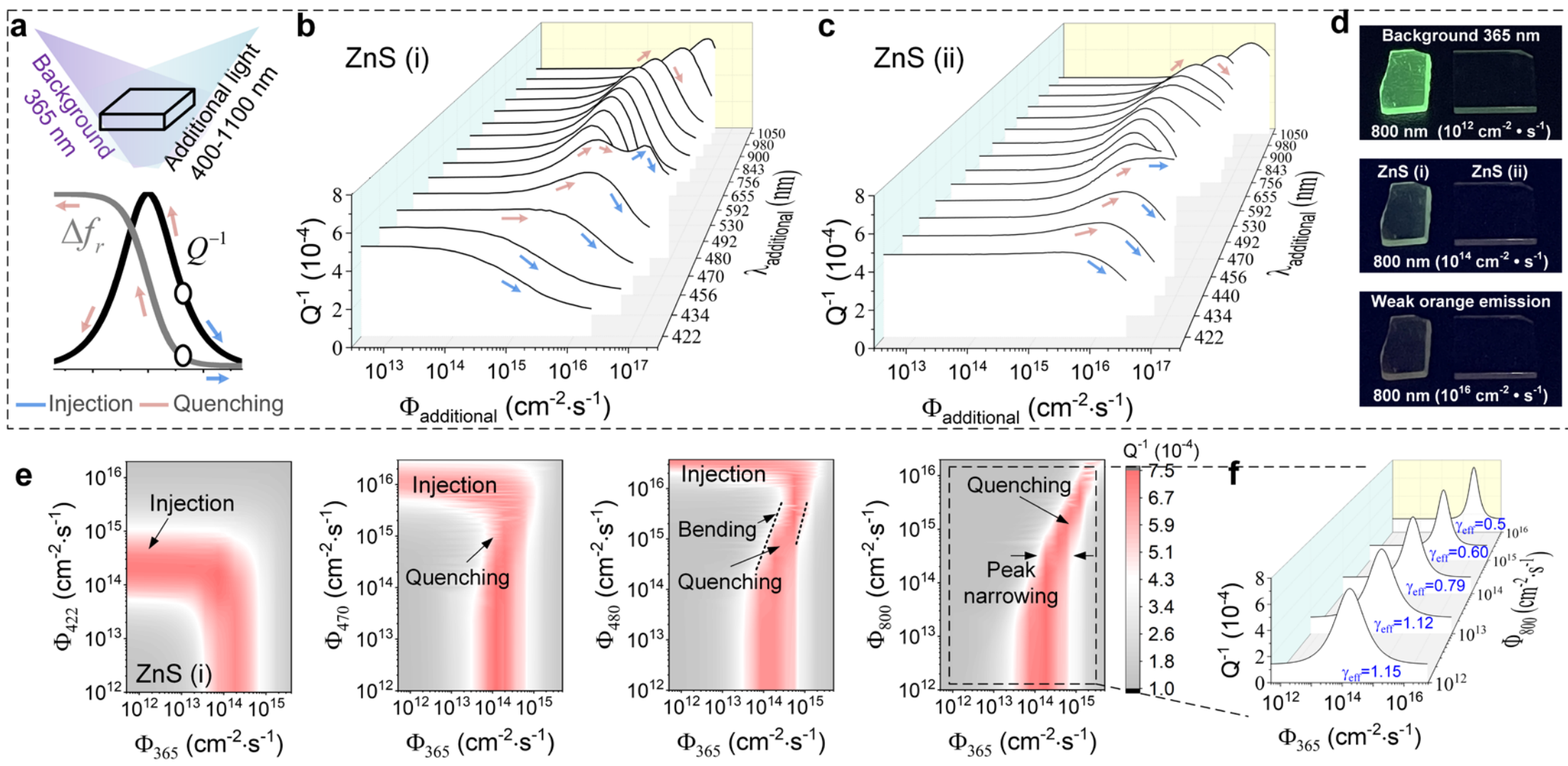

**Fig. 4: Responses of ZnS to steady-state bichromatic light. a-d,** Damping and photoluminescence of ZnS (i) and (ii) under a constant-intensity 365 nm UV light combined with an additional light source (400-1100 nm) of varying photon flux $\Phi_{additional}$ ($10^{12}\ cm^{-2}\cdot s^{-1}$-$10^{17}\ cm^{-2}\cdot s^{-1}$). The restoration of $Q^{-1}$ and the weakening of green emission is defined as the quenching effect (indicated by the red arrow), while the opposite process is referred to as the injection effect (blue arrow). **e and f,** Damping contour maps and damping waterfall plot of ZnS (i) under different bichromatic illumination conditions.

### 3.3 The coupling of recombination pathways under transient illumination

Under steady-state illumination, recombination pathways in ZnS are strongly coupled, and $r_{total}$ is well described by the multiplicative weighting model ($p$→0 limit of power-mean model). To probe the transient regime, we measured the time-dependent $Q^{-1}$ of ZnS (i) during abrupt on and off switching of monochromatic light (320, 365, 456, and 470 nm) and bichromatic light (365 nm + 800 nm; the 800 nm remained on while only the 365 nm was toggled; $\Phi_{800} = 10^{16} cm^{-2}\cdot s^{-1}$) (Fig. 5**a**, **b**). $Q^{-1}$ remains constant in initial and final states, ensuring a consistent carrier concentration for each condition. The equilibration time of $Q^{-1}$ provides a qualitative measure of recombination kinetics, with faster equilibration indicating a higher total recombination rate $r_{total}$. At 320 nm, $r_{total}$ is slightly higher than at 365 nm due to the increased $w_{direct}$, consistent with the damping peak shift in the intrinsic absorption region under steady-state illumination (Fig. 3**c**). Likewise, at 470 nm and 365 nm + 800 nm, $r_{total}$ is markedly larger than at 365 nm due to the increased $w_{sub}$, again consistent with steady-state trends (Fig. 3**c** and 4**f**). These results suggest that pathway coupling also affects the transient regime.

To further examine whether the transient response can be captured by the multiplicative weighting model with fixed $w_i$, we measured switch-on and switch-off curves at 365 nm with varying photon flux ($\Phi$) (hollow squares in Fig. 5**c** and **d**) and compared with numerical solutions (solid lines) derived from the rate equations, $\frac{d\Delta n}{dt} = -c_{total} \cdot \Delta n^{\gamma_{eff}} + \alpha\beta\Phi$, where $\gamma_{eff} = 1.2$ and $c_{total} = 3.83 \times 10^{-4}\ cm^{0.6} \cdot s^{-1}$ were used (Supplementary Note 12 [35]). The agreement between numerical predictions and experimental data for the switch-on process further supports the multiplicative model. However, during switch-off, the numerical model predicts a substantially slower decay than observed experimentally. The corresponding time-dependent light emission also decays more rapidly during switch-off than it rises during switch-on. These discrepancies indicate that the weights $w_i$ may evolve dynamically during switch-off conditions, or that $r_{total}$ no longer strictly follows the multiplicative model, with $p$ no longer close to zero in the power-mean framework. This asymmetry likely arises because carrier generation is removed instantaneously during switch-off, whereas defect occupancy and carrier redistribution may relax on different timescales.

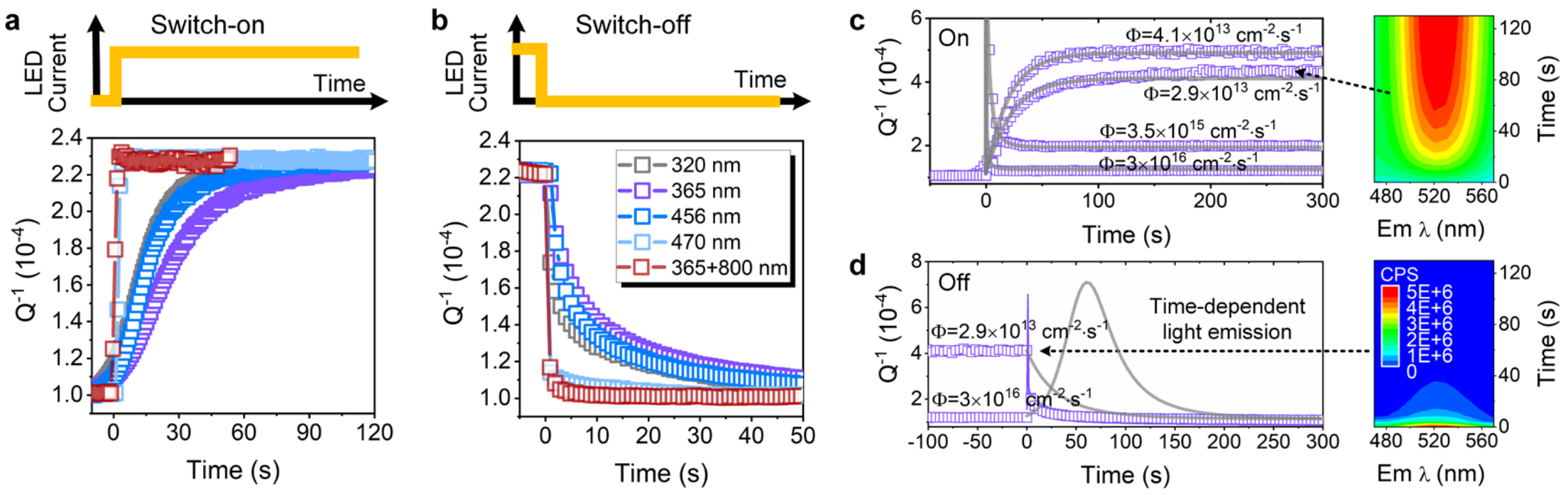

**Fig. 5:** Responses of ZnS (i) to transient illumination. **a and b,** Time dependence of $Q^{-1}$ during abrupt switch-on and switch-off of monochromatic and bichromatic light. **c and d,** Measured and numerically calculated (assuming constant $w_i$) time dependence of $Q^{-1}$ during switch-on and switch-off of 365 nm light at various photon fluxes ($\Phi$). Time-dependent light emissions were also measured.

### 3.4 Geometric representation of coupled pathways and carrier dynamics

We present a curved surface in cylindrical coordinates to visualize the multiplicative weighting model (Fig. 6**a**). This geometric representation should be viewed as a phenomenological map of the effective recombination exponent rather than a real-space trajectory. Here, $r$ denotes the total recombination rate, $\theta$ determines the weighting factors, and $z$ represents the carrier concentration $\Delta n$. As the photon energy decreases slightly below $E_g$ (from 320 to 365 nm), $\theta$ varies from nearly 0 to nearly π/2, leading to an increase in $w_{trap}$ and a decrease in $w_{direct}$. Correspondingly, the $r$-$\Delta n$ relationship shifts from a parabola to a linear form, reflecting the continuous transition from direct to trap-assisted recombination. When the photon energy of monochromatic light decreases from $E_g$ to $0.66E_g$-$0.8E_g$, or with the addition of long-wavelength light ($E_{photon}<0.8E_g$), $\theta$ shifts further from nearly $\pi/2$ to $\pi$, increasing $w_{sub}$ while decreasing $w_{trap}$. As a result, the $r$-$\Delta n$ relationship transitions from a linear trend to a root curve. During switch-on, the trajectory follows an approximately constant-$\theta$ path (Fig. 6**a**), indicating nearly fixed pathway weights $w_i$. In contrast, the switch-off process may deviate from this constant-$\theta$ path, or even leave the $p \to 0$ surface, suggesting dynamic evolution of $w_i$ or a departure from the multiplicative limit after carrier generation is removed.

Fig. 6**b-d** illustrate the carrier dynamics underlying the coupling between trap-assisted and sublinear recombination in ZnS (i) under steady-state illumination. Under 365 nm light, recombination center 1 ($RC_1$, identified as the zinc antisite through DFT calculations, Supplementary Note 13 [35]) facilitates carrier generation, trap-assisted recombination, and green light emission, whereas the other defect level, $RC_2$ (sulfur vacancy), remains fully occupied and does not participate in the recombination process (Fig. 6**b**). Introducing additional light in the 365-440 nm range increases $\Delta \mathrm{n}$ and $\Delta \mathrm{p}$ and enhances green light emission, without altering the weighting of the recombination pathway. However, illumination in the 440-520 nm range triggers electron redistribution between $RC_1$ and $RC_2$. Since $RC_2$ is more abundant than $RC_1$, electrons gradually saturate both centers. This process reduces the weighting of trap-assisted recombination, initiates sublinear recombination (Fig. 6**c**), reduces green emission, and induces weak orange emission. As both recombination centers become activated, $r_{total}$ increases, initially reducing $\Delta n$ (Fig. 6**c**, left). However, because 440-520 nm light also contributes to carrier generation, increasing photon flux eventually raises $\Delta n$ (Fig. 6**c**, right), resulting in peaks in both the $Q^{-1}$ and $\Delta f_r$ curves (Fig. 4). Beyond 520 nm, electron redistribution between $RC_1$ and $RC_2$ continues, but in the absence of additional carrier generation, $\Delta \mathrm{n}$ steadily decreases (Fig. 6**d**).

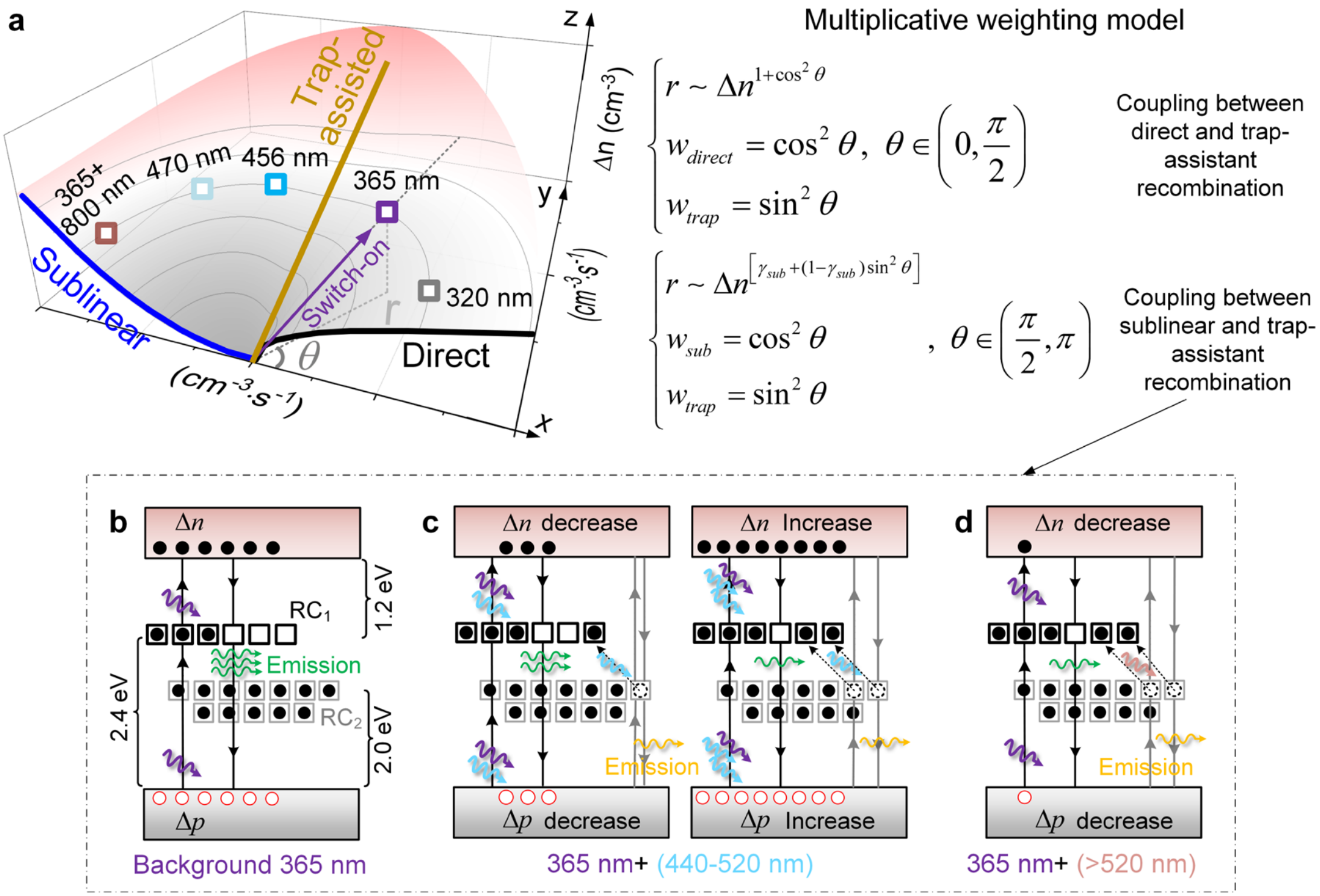

**Fig. 6: Geometric representation of coupled recombination pathways in ZnS and the associated carrier dynamics. a,** Multiplicative weighting model represented as a curved surface in cylindrical coordinates, illustrating coupling between recombination pathways. Here, $r$, $\theta$, and $z$ denote the total recombination rate, weighting factors, and carrier concentration Δn, respectively. For $\theta \in [0, \pi/2]$, the geometric model describes the coupling between direct and trap-assisted recombination; For $\theta \in [\pi/2, \pi]$, it represents coupling between sublinear and trap-assisted recombination. For the light switch-on, $w_i$ remains fixed, corresponding to a constant-$\theta$ trajectory. During switch-off, $w_i$ may evolve dynamically or the trajectory may leave the $p \to 0$ surface (not plotted here). **b-d,** Schematics of carrier dynamics underlying the coupling between trap-assisted and sublinear recombination in ZnS (i).

## 4. Conclusions

We developed light-induced mechanical absorption spectroscopy (LIMAS) to study recombination dynamics in semiconductors using mechanical damping. Our results on ZnS demonstrate that the total recombination rate under illumination follows a multiplicative weighting model, $r_{total} \propto \prod {r_i}^{w_i}$ with $\sum w_i = 1$. This multiplicative model corresponds to the $p \to 0$ limit of a generalized power-mean recombination model, $r_{total} \propto \left(\sum w_i r_i^p\right)^{1/p}$, indicating strong coupling between pathways. In contrast, previous work that relied on the additive model (the $p$=1 limit), which treats pathways as decoupled, may have overlooked some information and may be fruitfully revisited within the power-mean framework. While LIMAS is primarily applicable to piezoelectric semiconductors, it can serve as a valuable complement to traditional methods, and their combined use offers a promising route for exploring recombination mechanisms in materials with complex spectral line shapes.

## Data availability

The data that support the findings of this study are available from the corresponding author upon reasonable request.

## Acknowledgements

We acknowledge the financial support from the Discovery Grants Program of the Natural Sciences and Engineering Research Council of Canada (NSERC) RGPIN-2018-05731, the Ontario Early Researcher Award, and the Canada Foundation for Innovation (CFI) - Evans Leaders Fund (JELF) Number 38044.

## Author contributions

**Mingyu Xie:** Conceptualization, Methodology, Investigation, Data curation, Writing–original draft, Writing–review & editing. **Ruitian Chen:** Conceptualization, Methodology, Investigation, Writing–review & editing. **Jiaze Wu:** Methodology, Investigation, Writing–review & editing. **Kaiqi Qiu:** Investigation, Writing–review & editing. **Mingqiang Li:** Investigation, Writing–review & editing. **Huicong Chen:** Investigation, Writing–review & editing. **Kai Huang:** Conceptualization, Supervision, Writing–review & editing. **Yu Zou:** Conceptualization, Funding acquisition, Supervision, Writing–review & editing.

## Competing interests

There is no conflict of interest to declare.

# Supplementary Information

## Unveiling coupling between carrier recombination pathways in semiconductors via mechanical damping

Mingyu Xie[1], Ruitian Chen[1], Jiaze Wu[1], Kaiqi Qiu[1], Mingqiang Li[1], Huicong Chen[1], Kai Huang[1], Yu Zou[1,a)]
[1]Department of Materials Science and Engineering, University of Toronto, Toronto, ON M5S 3E4, Canada

[a)] Author to whom all correspondence should be addressed, Email: mse.zou@utoronto.ca

## Supplementary Note 1. Annotation of symbols

**Table 1.1 Annotation of symbols**

| Symbol | Expressions | Annotations |
|---|---|---|
| $c$ | | Elastic coefficient |
| $c_c$, $c_v$ | | Trap capture coefficients for electrons from conduction and valence bands |
| $c_D$, $c_A$, $c_T$, $c_S$ | | Coefficient for direct, Auger, trap-assisted and sublinear recombination |
| $D$ | | Electric displacement |
| $D_n$, $D_p$ | | Diffusion coefficients of electrons and holes |
| $D_{eq}$ | $\frac{D_n + \eta D_p}{1 + \eta}$ | Equivalent diffusion coefficient |
| $E$ | | Electric field |
| $E_F$ | | Fermi energy level |
| $e$ | | Piezoelectric coefficient |
| $e_c$, $e_v$ | | Trap emission coefficients to conduction and valence bands |
| $f$ | $\frac{\Delta n' - \Delta p'}{n_{net}}$ | The fraction of net space charge contributed by mobile carriers |
| $f_r$ | | Mechanical resonant frequency |
| $g$ | $\alpha\beta\Phi$ | Photogeneration rate |
| $i$ | $\sqrt{-1}$ | Imaginary unit |
| $J_n$, $J_p$ | | Current in conduction and valence bands |
| $k$ | $\omega/v_0$ | Elastic wavenumber |
| $N_c$, $N_v$ | | Effective density of states for conduction and valence bands |
| $N_r$ | | Concentration of the recombination center for trap-assisted recombination |
| $n_r$ | | Concentration of traps occupied by electrons |
| $n_0$, $p_0$ | | Thermal equilibrium carrier concentration |
| $\Delta n$, $\Delta p$ | | Photo-generated excess carrier concentration |
| $\Delta n_0$, $\Delta n_\infty$ | | Initial and final photo-generated excess carrier concentration |
| $\Delta n'$, $\Delta p'$ | | Perturbation of carrier concentrations induced by elastic vibration |
| $n_{net}$ | $-\frac{\nabla \cdot D}{q}$ | Net space charge concentration |
| $Q^{-1}$ | | Mechanical damping |
| $q$ | | Elementary charge |
| $r_i$, $r_{total}$ | | Individual and total recombination rate |
| $S$, $T$ | | Strain and stress |
| $t$, $u$ | | Time and displacement |
| $v_0$ | $\sqrt{c/\rho}$ | Elastic wave velocity |
| $w_i$ | | Weighting factor of each pathway ($w_{direct}$, $w_{trap}$, and $w_{sub}$ in this work) |
| $\alpha$, $\beta$ | | Absorption coefficient and quantum yield |
| $\gamma$ | | Index, $\gamma$=3,2,1,<1 for Auger, direct, trap-assisted and sublinear recombination |
| $\varepsilon$ | | Dielectric coefficient |
| $\eta$ | $-\Delta p'/\Delta n'$ | The ratio of $\Delta p'$ to $\Delta n'$ |
| $\lambda$ | | The wavelength of light |
| $\mu_n, \mu_p$ | | Electron and hole mobility |
| $\rho$ | | Mass density |
| $\sigma$ | $q(\mu_n n + \mu_p p)$ | Conductivity |
| $\tau_n, \tau_p$ | | Lifetimes of electrons and holes |
| $\tau_{eq}$ | $\frac{1+\eta}{\frac{\eta}{\tau_p}+\frac{1}{\tau_n}}$ | Equivalent lifetime |
| $\tau_T$ | $\frac{c_c+e_v}{c_c e_v N_r}$ | Lifetime of trap-assisted recombination |
| $\Phi$ | | Photon flux |
| $\omega, \omega_c, \omega_D, \omega_\tau$ | | Mechanical, conductivity, diffusion, and lifetime frequency |

# Supplementary Note 2. The setup of LIMAS

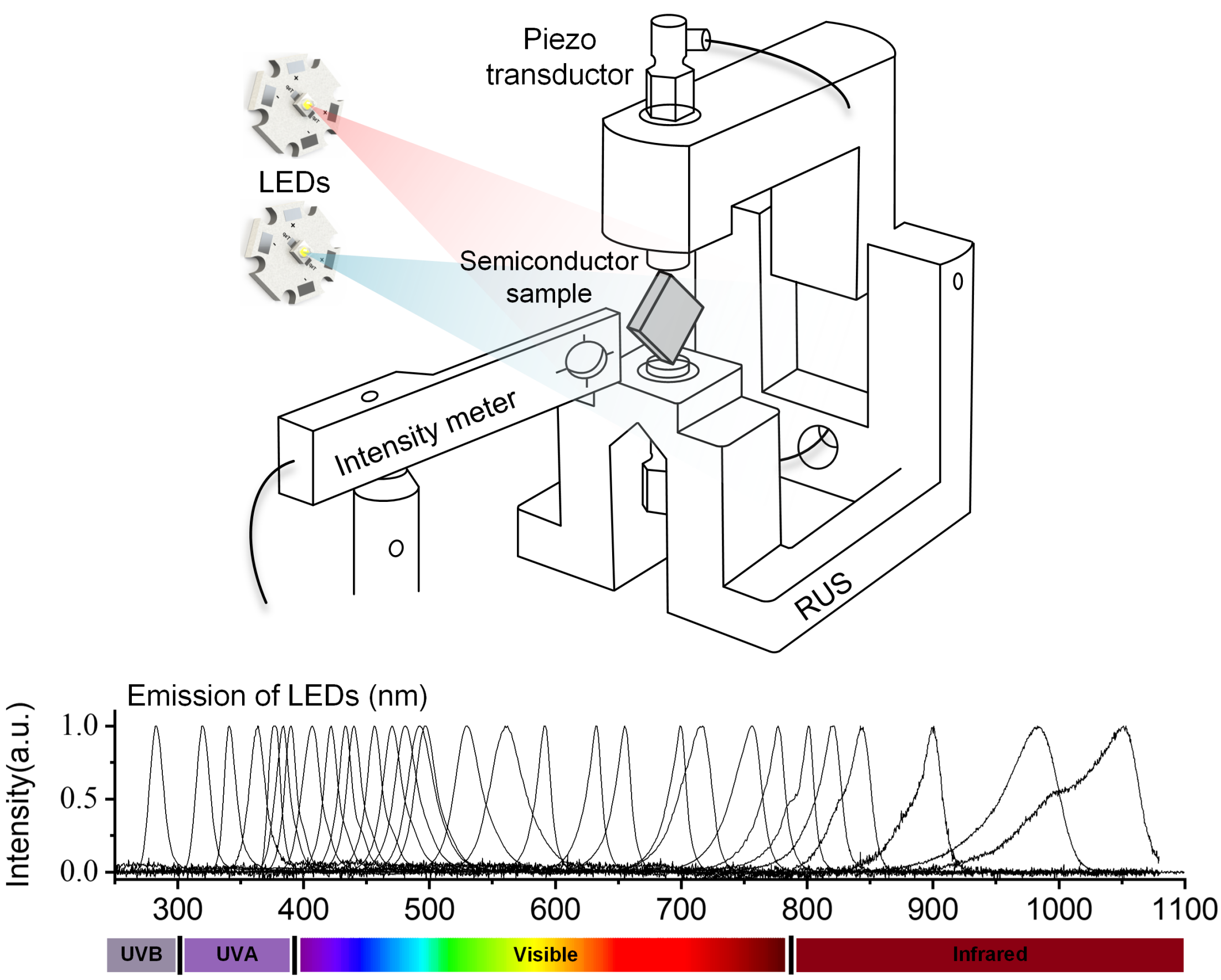

Supplementary Figure 2.1. Setup of light-induced mechanical absorption spectroscopy (LIMAS).

Figure 2.1 illustrates the setup of light-induced mechanical absorption spectroscopy (LIMAS) in this work. The system utilizes resonant ultrasound spectroscopy[1, 2] (RUS 008, Alamo Creek Engineering) to measure damping ( $Q^{-1}$ ) and resonance frequency changes ( $\Delta f_r$ ). The semiconductor sample is clamped between two opposite corners of two PZT transducers, which are mounted on a vibration isolation stage to minimize external disturbances. The transmitting PZT transducer generates elastic waves with a constant amplitude across a frequency range from kHz to MHz. The receiving PZT transducer detects a large peak in the amplitude-frequency curve when the excitation frequency matches one of the sample's eigenfrequencies (Supplementary Note 3). The $Q^{-1}$ value is derived from the half-bandwidth (-3 dB) of the peak, while $f_r$ is determined from the peak position (see Supplementary Note 7 for details). For illumination, various LEDs with emission wavelengths ranging from 300 nm to 1100 nm are used. These LEDs were calibrated using a spectrofluorometer (Horiba, FluoroMax Plus), and their corresponding emission spectra are presented below. The photon flux is measured using a Si-based photodiode power sensor (S120VC, 200 nm-1100 nm, Thorlabs) with a resolution of 1 nW. During the test, the photodiode power sensor is placed near the sample, while the LED is positioned along the central axis shared by both the sensor and the sample, at a sufficient distance from them, ensuring that the measured photon flux accurately represents the actual photon flux incident on the sample. A LabVIEW program is employed to control all the components of the setup. Experiments were conducted under both steady-state (monochromatic and bichromatic) and transient conditions (with a time resolution of 1 s in this work).

The current LIMAS implementation has several practical limitations. First, using discrete LEDs makes wavelength changes time-consuming, and the divergence of the LED emission limits the maximum photon flux incident on the sample. Second, when the photon energy of an LED exceeds the sample bandgap, carrier generation occurs only within a few optical penetration depths of the surface, potentially affecting the absolute value of $Q^{-1}$. Third, the sample size cannot be less than approximately $1\times1\times1$ mm$^3$, limiting this method to bulk crystals rather than microscale samples.

Aside from these practical concerns, LIMAS's main physical limitation is that it is only applicable to semiconductors with an effective piezoelectric response. Nevertheless, its scope could be broadened by employing different mechanical loading configurations. For example, integrating LIMAS with surface acoustic waves in the GHz range would enable ultrafast time-resolved measurements (such as hot-carrier relaxation) in semiconductor films (Figure 2.2**a**). Adopting contact-resonance damping detection, such as atomic force microscopy [3], could greatly enhance the spatial resolution of LIMAS and allow carrier recombination to be probed at the level of individual defects such as dislocations (Figure 2.2**b**). Exploring these extensions is an important direction of our future work.

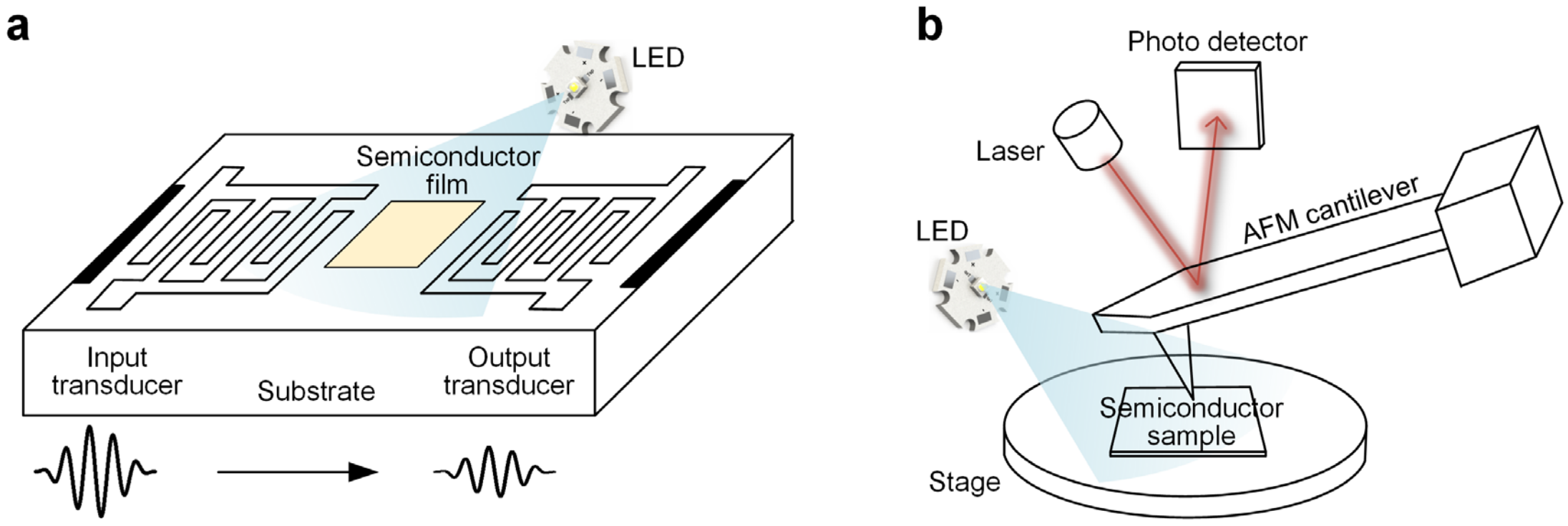

Supplementary Figure 2.2 Possible implementations of LIMAS with different mechanical loading configurations. a Surface acoustic wave-based configuration. b Contact-resonance AFM based configuration

## Supplementary Note 3. Piezoelectrically active/inactive eigenmodes

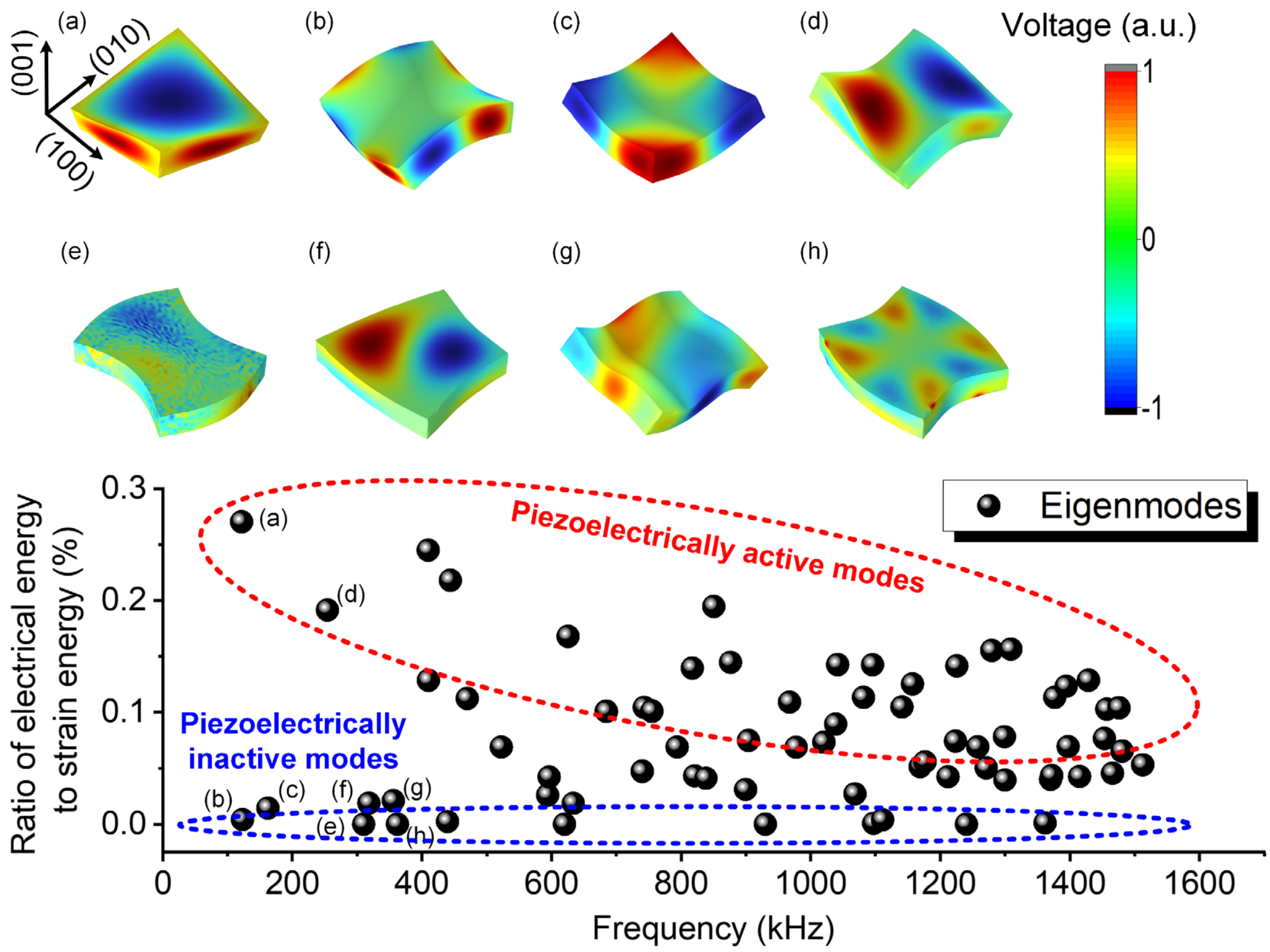

Supplementary Figure 3.1. Piezoelectrically active and inactive eigenmodes of a 5×5×1 mm ZnS wafer.

Figure 3.1 illustrates the finite element method (FEM) calculated first eight eigenmodes of a 5×5×1 mm ZnS wafer, along with the first ninety ratios of electrical energy to strain energy. A "piezoelectrically inactive eigenmode" is defined when the ratio approaches zero (blue circle), whereas a "piezoelectrically active eigenmode" corresponds to a nonzero ratio (red circle). For the first eight eigenmodes, (a) and (d) belong to the "piezoelectrically active eigenmode", while (e) and (h) belong to the "piezoelectrically inactive eigenmode". In LIMAS, only piezoelectrically active eigenmodes should be used to analyze carrier recombination processes to ensure a high signal-to-noise ratio (SNR) during testing.

The parameters of ZnS used for FEM calculation here are as follows: mass density 4088 kg/m$^3$, elastic coefficients c11=c22=c33=104.6 GPa, c12=c13=c23=65.3 GPa, c44=c55=c66=46.13 GPa. Piezoelectric coefficients e14=e24=e36=0.14 C/m$^2$. Relative dielectric coefficients $\varepsilon_{11} = \varepsilon_{22} = \varepsilon_{33} = 8.3$.

## Supplementary Note 4. Derivation of Eq. (1)

For compound semiconductors, the constitutive law is[4]:

$$\begin{cases} T = \mathrm{c}S - eE \\ D = eS + \varepsilon E \end{cases} \tag{S1}$$

where $T$, $S$ are stress and strain; $D$, $E$ are electric displacement and electric field. $c$, $e$, and $\varepsilon$ are the elastic, piezoelectric and dielectric constants, respectively. When a uniformly illuminated sample undergoes mechanical vibration, an internal alternating electric field ($E$) is generated, driving carriers to move and redistributing electrons in localized states (Fig. 2**c**). Under this circumstance, $n = n_0 + \Delta n + \Delta n', p = p_0 + \Delta p + \Delta p', \Delta p' = -\eta\Delta n', \nabla \cdot D = -qn_{net}$. It is reasonable to assume that the ratio of $\Delta p'$ to $\Delta n'$ is a constant, $\eta \in (0, \infty)$, depending on the specific material and doping conditions. $n_{net}$ denotes net space charge concentration. $fn_{net} = (\Delta n' - \Delta p')$, which means a fraction $f$ of the net space charge is caused by mobile carriers, and the left is due to the redistribution of electrons in localized states (Fig. 2**c**). Considering the 1D problem**,** the vibration, continuity, and transport equations can be written as:

$$\begin{cases} \dfrac{\partial T}{\partial x} = \rho \dfrac{\partial^2 u}{\partial t^2} \\ \dfrac{\partial(\Delta n + \Delta n')}{\partial t} = \dfrac{1}{q}\dfrac{\partial J_n}{\partial x} - \dfrac{\Delta n + \Delta n'}{\tau_n} + g, \quad \dfrac{\partial(\Delta p + \Delta p')}{\partial t} = -\dfrac{1}{q}\dfrac{\partial J_p}{\partial x} - \dfrac{\Delta p + \Delta p'}{\tau_p} + g \\ J_n = qE\mu_n n + qD_n \dfrac{\partial \Delta n'}{\partial x}, \qquad J_p = qE\mu_p p - qD_p \dfrac{\partial \Delta p'}{\partial x} \end{cases} \tag{S2}$$

where $u, \rho$ are displacement and mass density. $J_n$, $J_p$ are current in the conduction and valence band. $\tau_n$, $\tau_p$ are lifetimes of electrons and holes. $g = \alpha\beta\Phi$ is the photogeneration rate, where $\alpha$ is the absorption coefficient, $\beta$ is quantum yield (carrier pairs generated by each photon), $\Phi$ is the photon flux. $\mu_n$, $\mu_p, D_n, D_p$ are mobility and diffusion coefficients of electrons and holes. During test, the mechanical vibration amplitude in the sample is negligibly small, ensuring no impact on carrier generation or recombination, so $\frac{d\Delta n}{dt} = -\frac{\Delta n}{\tau_n} + g$, $\frac{d\Delta p}{dt} = -\frac{\Delta p}{\tau_p} + g$, and

$$\frac{\partial(\Delta n' - \Delta p')}{\partial t} = \frac{1}{q}\frac{\partial(J_n + J_p)}{\partial x} + \frac{\Delta p'}{\tau_p} - \frac{\Delta n'}{\tau_n} \tag{S3}$$

where $\frac{1}{q}\frac{\partial(J_n+J_p)}{\partial x} = \frac{\partial E}{\partial x}(\mu_n n + \mu_p p) + E\mu_n \frac{\partial \Delta n'}{\partial x} + E\mu_p \frac{\partial \Delta p'}{\partial x} + D_n \frac{\partial^2 \Delta n'}{\partial x^2} - D_p \frac{\partial^2 \Delta p'}{\partial x^2}$, $\Delta p' = -\eta \Delta n'$, and $(\Delta n' - \Delta p') = -\frac{f}{q}\nabla \cdot D$. For small signals, the product of the differentials of $D$ and $E$ is negligible, so

$$\frac{\partial^2 D}{\partial x \partial t} = -\frac{\sigma}{f}\frac{\partial E}{\partial x} + D_{eq}\frac{\partial^3 D}{\partial x^3} - \frac{1}{\tau_{eq}}\frac{\partial D}{\partial x} \tag{S4}$$

Where $\sigma = q[\mu_n(n_0 + \Delta n) + \mu_p(p_0 + \Delta p)]$ is the conductivity, $D_{eq} = (D_n + \eta D_p)/(1+\eta)$ is the equivalent diffusion coefficient, $\tau_{eq} = (1+\eta)/\left(\frac{\eta}{\tau_p} + \frac{1}{\tau_n}\right)$ is the equivalent lifetime. As $\eta$ approaches 0, conduction band electrons dominate, leading to $D_{eq} = D_n$ and $\tau_{eq} = \tau_n$. Conversely, as $\eta$ approaches $\infty$, valence band holes dominate, resulting in $D_{eq} = D_p$ and $\tau_{eq} = \tau_p$.

Assume the time and space dependencies of the plane elastic wave as: $D = D_0 e^{i(kx-\omega t)}$, $E = E_0 e^{i(kx-\omega t)}$. Where $k = \omega/v_0$ is the wave number, $v_0 = \sqrt{c/\rho}$ is the wave velocity, $\omega$ is circular frequency of the mechanical vibration, which is typically $10^5$-$10^7$ rad/s in our tests. By first solving Eq. (S4) and then substituting the obtained *D*-*E* relationship into Eq. (S1), we can determine the mechanical damping ($Q^{-1}$) and resonance frequency change ($\Delta f_r/f_r$):

$$\begin{cases} Q^{-1} = \dfrac{e^2}{c\varepsilon} \dfrac{\omega_c / \omega}{1 + \left(\omega_c / \omega + \omega / \omega_D + \omega_\tau / \omega\right)^2} \\ \dfrac{\Delta f_r}{f_r} = -\dfrac{e^2}{2c\varepsilon} \left[ 1 - \dfrac{1 + \left(\omega / \omega_D + \omega_\tau / \omega\right)\left(\omega / \omega_D + \omega_\tau / \omega + \omega_c / \omega\right)}{1 + \left(\omega_c / \omega + \omega / \omega_D + \omega_\tau / \omega\right)^2} \right] \end{cases} \tag{S5}$$

Where $\omega_c = \sigma/(f\varepsilon)$ is the conductivity frequency, $\omega_D = v_0^2/D_{eq}$ is the diffusion frequency, $\omega_\tau = 1/\tau_{eq}$ is the lifetime frequency.

The response of $Q^{-1}$ and $\Delta f_r/f_r$ to illumination can vary depending on the semiconductor type and test conditions (frequency and temperature). Below, we discuss two scenarios:

## Case I: Strong *n*-type or *p*-type compound semiconductors at room temperature

For strong *n*-type semiconductors at room temperature (the case for *p*-type is similar), impurity ionization is significant, resulting in a high electron concentration in the conduction band ($n_0$) compared to the hole concentration in the valence band ($p_0$). Consequently, the perturbation of electrons induced by mechanical vibration is much larger than that of holes, leading to $\eta \approx 0$. The corresponding parameters are: $\sigma = q\mu_n(n_0 + \Delta n)$, $\omega_D = v_0^2/D_n$, $\omega_\tau = 1/\tau_n$. If there is no illumination, then $\Delta n = 0$. Table 4.1 lists the $\omega_c$, $\omega_D$, and $\omega_\tau$ values of several commercial strong *n*-type and *p*-type semiconductors in darkness. For direct bandgap *n*-type or *p*-type semiconductors, carrier lifetime is estimated based on direct recombination, thus $\omega_\tau \approx c_D n_0$, where $c_D$=10^-9^$cm^3/s$ is the estimated direct recombination coefficient. $\mu_n$, $D_n$ and $\mu_p, D_p$ follow the Einstein relation.

**Table 4.1 Parameters of some commercial *n*/*p*-type semiconductors at room temperature[5]**

| Semi conductor | Conductivity type | $n_0/p_0$ ($cm^{-3}$) | $\mu_n/\mu_p$ ($cm^2/$V/s) | σ ($ohm^{-1}$ $cm^{-1}$) | $\varepsilon_r$ | $\omega_c$ (rad/s) | $v_0^2$ ($cm^2/s^2$) | $D_n/D_p$ ($cm^2/$s) | $\omega_D$ (rad/s) | $\omega_\tau$ (rad/s) |
|---|---|---|---|---|---|---|---|---|---|---|
| GaAs-Zn | *p*-type | 2E19 | 70 | 224 | 13.1 | 1.9E14 | 2.2E11 | 1.8 | 1.2E11 | 2E10 |
| GaAs-Te | *n*-type | 5E17 | 3500 | 280 | 13.1 | 2.4e14 | 2.2E11 | 91 | 2.4E9 | 5E8 |
| GaP | $n$-type | 4E16 | 470 | 3 | 11.4 | 3.0E12 | 3.6E11 | 12.2 | 3E10 | / |
| GaP-S | $n$-type | 7E17 | 270 | 30 | 11.4 | 3e13 | 3.6E11 | 7 | 5E10 | / |
| GaSb | $p$-type | 1.5E17 | 700 | 17 | 16.8 | 1.1E13 | 1E11 | 18.2 | 5.5E9 | 1.5E8 |
| GaSb-Zn | $p$-type | 3E18 | 400 | 192 | 16.8 | 1.3E14 | 1E11 | 10.4 | 9.6E9 | 3E9 |
| InAs | $n$-type | 3E16 | 2E4 | 96 | 12.3 | 8.8E13 | 1.6E11 | 520 | 3E8 | 3E7 |
| InAs-Zn | $p$-type | 5E18 | 130 | 104 | 12.3 | 9.5E13 | 1.6E11 | 3.4 | 4.7E10 | 5E9 |
| InP | $n$-type | 8E15 | 4000 | 5 | 9.6 | 6E12 | 1.2E11 | 104 | 1.1E9 | 8E6 |
| InP-Zn | $p$-type | 1E18 | 3000 | 480 | 9.6 | 5.6E14 | 1.2E11 | 78 | 1.5E9 | 1E9 |
| InSb-Te | $n$-type | 5E17 | 12500 | 1000 | 17.7 | 6.4E14 | 1.1E11 | 325 | 3.4E8 | 5E8 |
| InSb-Ge | $n$-type | 5E16 | 250 | 2 | 17.7 | 1.3E12 | 1.1E11 | 6.5 | 1.7E10 | 5E7 |

Here, we consider the mechanical vibration circular frequency ω is typically $10^5$-$10^7$ rad/s. According to Table 4.1, $\omega, \omega_\tau \ll \omega_D \ll \omega_c$, resulting in $Q^{-1}$ in Eq. (S5) being zero, while $\Delta f_r/f_r$ remains constant. $Q^{-1}$ and $\Delta f_r/f_r$ have no response to illumination in this condition.

## Case II: Compound semi-insulators at room temperature

In this case, impurity ionization is weak, so both conduction band electrons and valence band holes contribute to conductivity, i.e., $\sigma = \mathrm{q}\left[\mu_n(n_0 + \Delta n) + \mu_p(p_0 + \Delta p)\right]$. If there is no illumination, $\Delta\mathrm{n}, \Delta\mathrm{p} = 0$. Table 4.2 lists the values of $\omega_c$, $\omega_D$, and $\omega_\tau$ for several commercial semi-insulators at room temperature under darkness.

**Table 4.2 Parameters of some commercial semi-insulators at room temperature[5]**

| Semi-insulator | Conductivity type | $n_0/p_0$ ($cm^{-3}$) | $\mu_n/\mu_p$ ($cm^2$/V/s) | σ ($ohm^{-1}$ $cm^{-1}$) | $\varepsilon_r$ | $\omega_c$ (rad/s) | $v_0^2$ ($cm^2/s^2$) | $D_n/D_p$ ($cm^2$/s) | $\omega_D$ (rad/s) | $\omega_\tau$ (rad/s) |
|---|---|---|---|---|---|---|---|---|---|---|
| GaAs | S-I | / | 5500 | 1E-8 | 13.1 | 8.6E3 | 2.2E11 | 143 | 1.5E9 | 1E-2 |
| GaP | S-I | 9E8 | 150 | 1E-8 | 11.4 | 1E4 | 3.6E11 | 3.9 | 9.2E10 | 4E-1 |
| InP | S-I | / | 2500 | 2E-8 | 9.6 | 2.3E4 | 1.2E11 | 65 | 1.9E9 | 5E-2 |
| InGaAs-Fe | S-I | / | 1500 | 5E-8 | 13.9 | 4E4 | 1E11 | 40 | 2.6E9 | 2E-1 |
| CdS | S-I | / | 350 | 1E-10 | 6 | 188 | 3.4E11 | 9.1 | 3.7E10 | 1.8E-3 |
| CdSe | S-I | / | 1100 | 1E-11 | 6.5 | 17 | 9E10 | 28.6 | 3.2E9 | 5.7E-5 |
| CdTe | S-I | / | 1050 | 1E-11 | 9.0 | 12 | 9E10 | 27.3 | 3.3E9 | 6E-5 |
| ZnSe | S-I | / | 500 | 1E-8 | 6.6 | 1.7E4 | 1.2E11 | 13 | 9.2E9 | 1.2E-1 |
| ZnTe | S-I | / | 340 | 1E-8 | 9.6 | 1.1E4 | 9E10 | 8.84 | 1E10 | 1.8E-1 |
| ZnS | S-I | / | 500 | 1E-10 | 10 | 113 | 1.6E11 | 13 | 1.2E10 | 1.3E-3 |

According to Table 4.2, in the dark, $\omega_\tau \ll \omega_c < \omega \ll \omega_D$, so

$$\begin{cases} Q^{-1} = \dfrac{e^2}{c\varepsilon} \dfrac{\omega_c / \omega}{1+\left(\omega_c / \omega\right)^2} \\ \dfrac{\Delta f_r}{f_r} = -\dfrac{e^2}{2c\varepsilon} \dfrac{\left(\omega_c / \omega\right)^2}{1+\left(\omega_c / \omega\right)^2} \end{cases} \tag{S6}$$

For steady-state illumination, $\Delta n \gg n_0$, $\Delta p \gg p_0$, $\omega_c = q(\mu_n \Delta n + \mu_p \Delta p)/(f\varepsilon)$. When $\omega_c < \omega$, $Q^{-1}$ increases with photo flux $\Phi$, when $\omega_c = \omega$, $Q^{-1}$ reaches its maximum, and when $\omega_c > \omega$, $Q^{-1}$ decreases as photo flux increases. During illumination switching on/off, $\Delta n$ and $\Delta p$ become time-dependent, leading to time-dependent behavior in $Q^{-1}$ and $\Delta f_r / f_r$.

In this study, the ZnS single crystal samples fall under the second scenario, so Eq. (S6) is used.

## Supplementary Note 5. $\gamma$ value, damping, and resonance frequency change for each individual recombination pathway

According to Eq. (S6), $Q^{-1}$ and $\Delta f_r/f_r$ depend on $\omega_c$. $\omega_c = q(\mu_n\Delta n + \mu_p\Delta p)/(f\varepsilon)$, where $\Delta n$ and $\Delta p$ are functions of photo flux $\Phi$ for steady-state conditions or function of time *t* for transient process, determined by the overall recombination kinetics. Since the mechanical vibration amplitude is negligibly small, the carrier generation and recombination processes can still be described as:

$$\begin{cases} \dfrac{d\Delta n}{dt} = -\dfrac{\Delta n}{\tau_n} + g \\ \dfrac{d\Delta p}{dt} = -\dfrac{\Delta p}{\tau_p} + g \end{cases} \tag{S7}$$

Below, we consider different individual recombination pathways.

### Direct recombination

Considering direct recombination in semi-insulators (see process (i) in Fig. 2**d**):

$$r_{direct} = c_D\left(n_0\Delta p + p_0\Delta n + \Delta n\Delta p\right) \tag{S8}$$

where $c_D$ is direct recombination coefficient. Assuming $\Delta\mathrm{n}$ and $\Delta\mathrm{p}$ are significantly larger than the trap concentration, we obtain $\Delta n = \Delta p$, $\frac{\Delta n}{\tau_n} = \frac{\Delta p}{\tau_p} = c_D\Delta n\Delta p$ $(\Delta n, \Delta p \gg n_0, p_0)$. For steady-state illumination, $\frac{\Delta n}{\tau_n} = c_D(\Delta n)^2 = g = \alpha\beta\Phi$, $\omega_c = \frac{q(\mu_n+\mu_p)}{(f\varepsilon)}(\alpha\beta/c_D)^{1/2}\Phi^{1/2}$, so $\gamma = 2$.

For switch-off processes, $\frac{d\Delta n}{dt} = -c_D(\Delta n)^2$, $\Delta n = \frac{1}{c_D t + 1/\Delta n_0}$, where $\Delta n_0$ represents the initial photo-generated electron concentration. For switch-on, $\frac{d\Delta n}{dt} = -c_D(\Delta n)^2 + g$ , $\Delta n = \Delta n_\infty \tanh(\sqrt{\Delta n_\infty^2 c_D^2}t)$, where $\Delta n_\infty = (g/c_D)^{1/2}$ represents the final photo-generated electron concentration.

**Trap-assisted recombination (Shockley-Read-Hall recombination)**

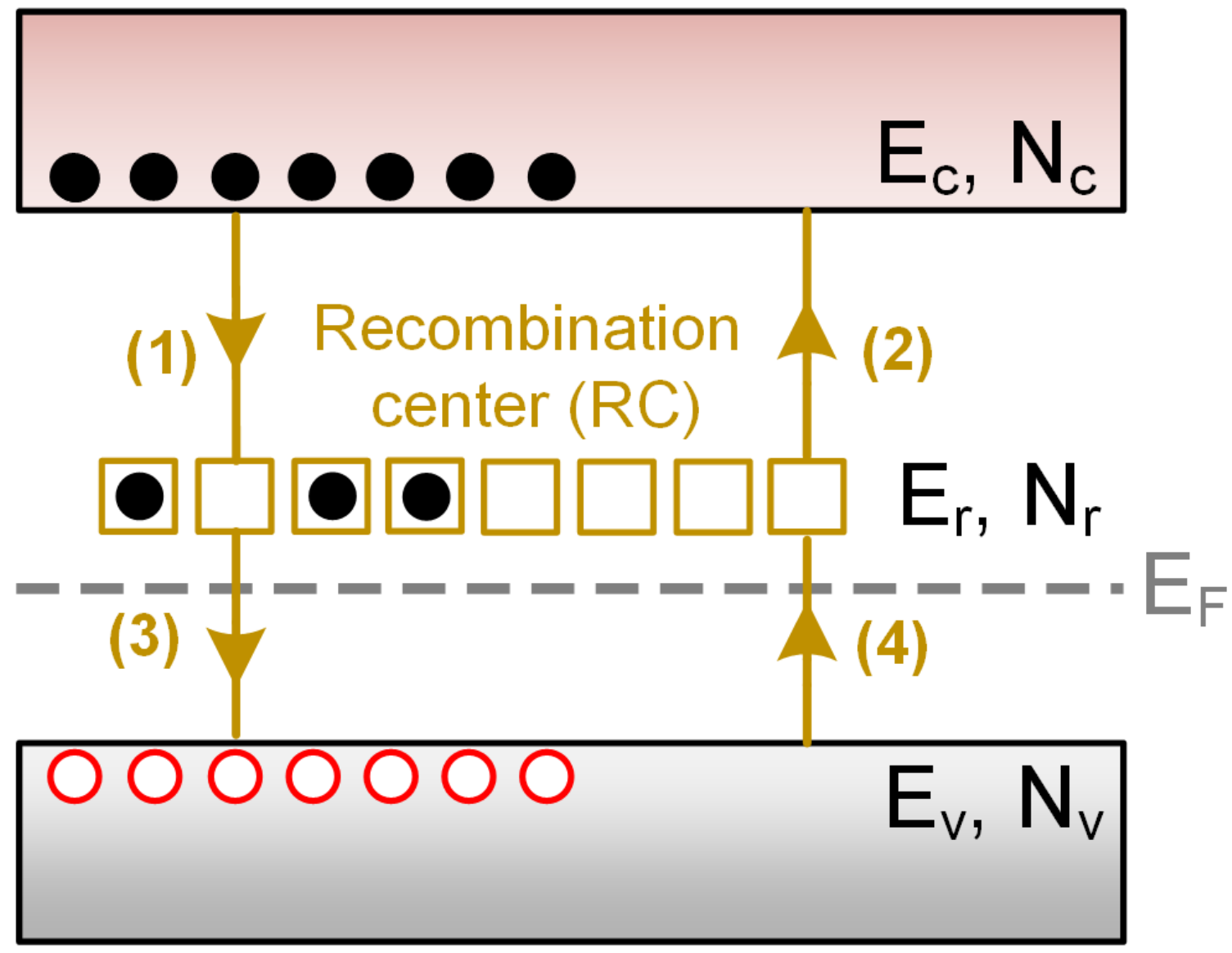

Supplementary Figure 5.1. Trap-assisted recombination, also known as Shockley-Read-Hall (SRH) recombination[6], occurring via a defect level within the bandgap.

Figure 5.1 shows the schematic of trap-assisted recombination. For processes (1) and (4), the capture rates are:

$$r_1 = c_c n\left(N_r - n_r\right) \tag{S9}$$

$$r_4 = c_v\left(N_r - n_r\right)\left(N_v - p\right) = c_v\left(N_r - n_r\right)N_v \tag{S10}$$

where $N_r$ is the trap density, i.e., the concentration of recombination center. $n_r$ is concentration of traps occupied by electrons. $c_c$ and $c_v$ are trap capture coefficients for electrons from the conduction and valence bands, respectively.

For processes (2) and (3), the emission rates are:

$$r_2 = e_c n_r\left(N_c - n\right) = e_c n_r N_c \tag{S11}$$

$$r_3 = e_v n_r p \tag{S12}$$

where $e_c$ and $e_v$ are the electron emission coefficients for conduction and valence bands.

Considering thermal equilibrium:

$$\begin{cases} c_c n_0 \left(N_r - n_{r0}\right) = e_c n_{r0} N_c \\ n_0 = N_c e^{-\frac{E_c - E_F}{k_B T}} \\ n_{r0} = N_\mathrm{r} \dfrac{1}{1 + e^{\frac{E_r - E_F}{k_B T}}} \end{cases} \tag{S13}$$

where $E_F$ is the Fermi level. So, we obtain:

$$e_c = c_c e^{-\frac{E_c - E_F}{k_B T}} \left(1 + e^{\frac{E_r - E_F}{k_B T}} - 1\right) = c_c e^{-\frac{E_c - E_r}{k_B T}} = c_c \frac{n_1}{N_c} \tag{S14}$$

and

$$c_v = \frac{e_v}{N_v} p_1 \tag{S15}$$

where $n_1$ and $p_1$ are the concentration of conduction band electrons and valence band holes, respectively, when the Fermi level coincides with the trap energy level.

Considering the steady-state:

$$c_c n\left(N_r - n_r\right) + \frac{e_v p_1}{N_v}\left(N_r - n_r\right) N_v = c_c \frac{n_1}{N_c} n_r N_c + e_v n_r p \tag{S16}$$

and

$$n_r = \frac{c_c n + e_v p_1}{c_c (n + n_1) + e_v (p + p_1)} N_r \tag{S17}$$

The recombination rate can be written as:

$$r_{trap} = r_1 - r_2 = r_3 - r_4 = \frac{c_c e_v \left(np - n_1 p_1\right)}{c_c \left(n + n_1\right) + e_v \left(p + p_1\right)} N_r = \frac{c_c e_v N_r}{c_c + e_v} \Delta n \tag{S18}$$

where $\Delta n = \Delta p \geq n_1, p_1$. $\frac{\Delta n}{\tau_n} = \frac{\Delta p}{\tau_p} = \frac{c_c e_v N_r}{c_c + e_v} \Delta n = \frac{\Delta n}{\tau_T} = c_T \Delta n$. $\tau_T$ is the lifetime for trap-assisted recombination. For steady-state, $\omega_c = \frac{q(\mu_n + \mu_p)}{(f\varepsilon)} \alpha \beta \tau_T \Phi$, $\gamma = 1$.

Considering the transient process:

$$\begin{cases} \dfrac{dn_r}{dt} = r_1 + r_4 - r_2 - r_3 \\ \dfrac{dn}{dt} = -(r_1 - r_2) + g \\ \dfrac{dp}{dt} = -(r_3 - r_4) + g \end{cases} \tag{S19}$$

Recombination depends on both free carriers and trap occupancy dynamics ($n_r(t)$), so $\Delta$n and $\Delta$p must be obtained by numerically solving the above coupled equations. Figure 5.2 compares these numerical solutions with the steady-state approximation, showing that at high photon fluxes (large carrier densities) these two approaches converge. In this condition, for switch-off, $\Delta n = \Delta n_0 e^{-t/\tau_T}$. For switch-on, $\Delta n = \Delta n_\infty (1 - e^{-t/\tau_T})$, $\Delta n_\infty = g\tau_T$.

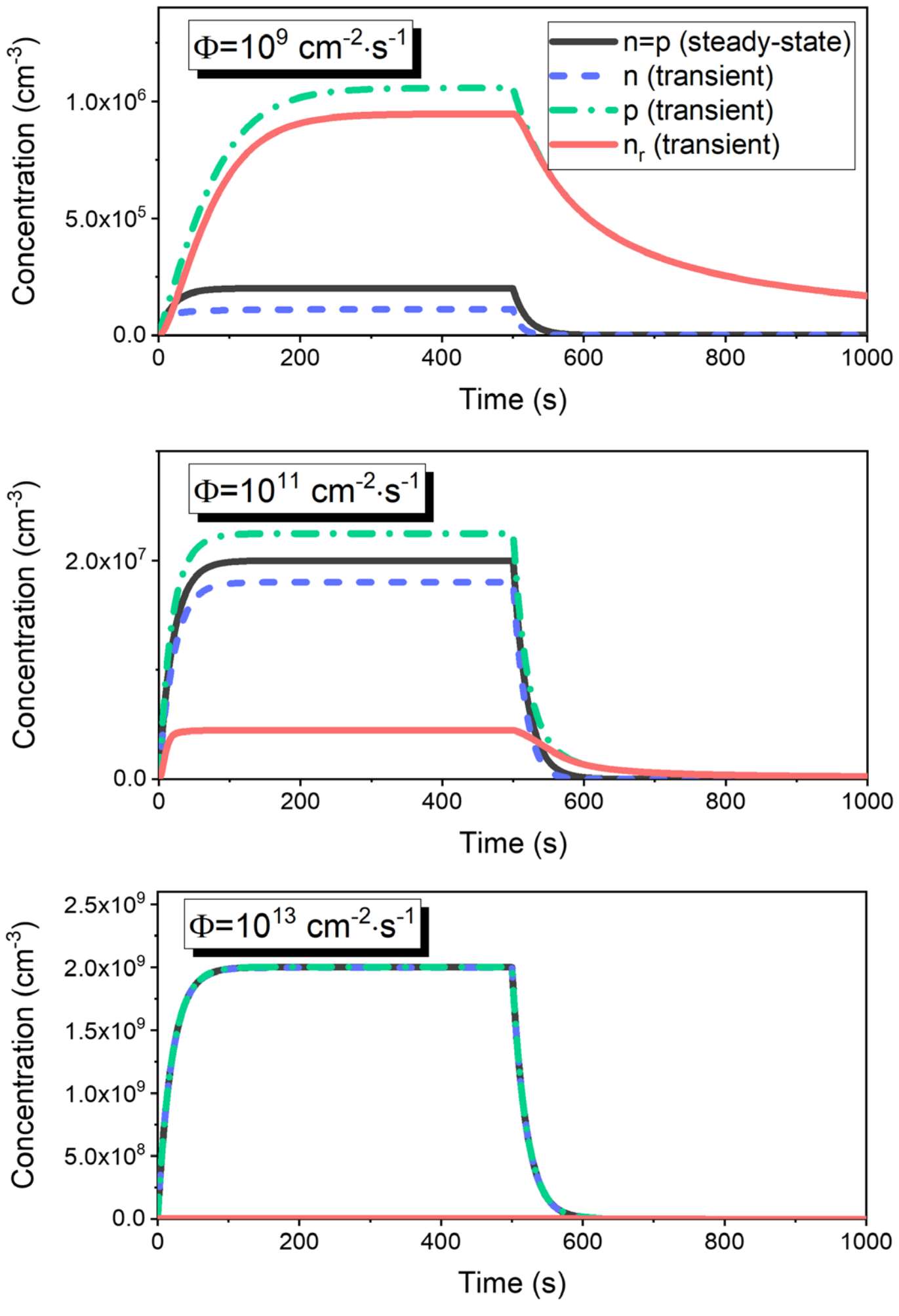

Supplementary Figure 5.2 Compares carrier concentrations obtained from numerical solutions of the full rate equations with those calculated using the steady-state recombination rate under different photon fluxes. The two approaches diverge at low fluxes due to trap dynamics but converge at high fluxes, where the steady-state approximation accurately captures the transient behavior.

## Auger recombination

$$r_{Auger} = c_A\left(n^3 - n_0^3\right) = c_A \Delta n^3 \tag{S20}$$

For Auger recombination[7] (see process (iii) in Fig. 2**d**), $\frac{\Delta n}{\tau_n} = \frac{\Delta p}{\tau_p} = c_A(\Delta n)^3$, where $c_A$ is the Auger recombination coefficient. Here, we still assume $\Delta n = \Delta p$. For steady-state, $\omega_c = \frac{q(\mu_n+\mu_p)}{(f\varepsilon)}(\alpha\beta/c_A)^{1/3}\Phi^{1/3}$, $\gamma = 3$. For switch-off, $\Delta n = \frac{1}{\sqrt{2c_A t + 1/(\Delta n_0)^2}}$. Auger recombination is ruled out in ZnS, as it primarily occurs in heavily doped semiconductors.

## Sublinear recombination

$$r_{sublinear} = c_\gamma \Delta n^\gamma, \quad \gamma \in (0,1) \tag{S21}$$

Sublinear recombination[8, 9] is more complex than trap-assisted recombination, as it involves multiple saturated defect levels (see process (iv) in Fig. 2**d**). Due to the saturation of these defect levels, the recombination rate is not proportional to the carrier concentration, resulting in a recombination exponent $\gamma$ < 1. In general case, $\frac{\Delta n}{\tau_n} = \frac{\Delta p}{\tau_p} = c_\gamma(\Delta n)^\gamma$, $\gamma$=3, 2, and 1 represent the above-mentioned Auger, direct, and trap-assisted recombination, respectively. When $\gamma$ <1, representing the sublinear recombination. For steady-state sublinear recombination, $\omega_c = \frac{q(\mu_n+\mu_p)}{(f\varepsilon)}\left(\alpha\beta/c_\gamma\right)^{1/\gamma}\Phi^{1/\gamma}$. For switch-off, $\Delta n = (1-\gamma)^{\frac{1}{1-\gamma}}\left[\frac{1}{1-\gamma}(\Delta n_0)^{1-\gamma} - c_\gamma t\right]^{\frac{1}{1-\gamma}}$.

## Other recombination

Other types of recombination, such as surface recombination (primarily occurs at the surface or interface), exciton recombination (primarily occurs in quantum dots), and band-tail recombination (primarily occurs in amorphous), are not considered in this high-purity bulk ZnS single crystal here.

## Supplementary Note 6. Parameters used to calculate Fig. 2e

For calculation of Fig. 2**e**, the mechanical vibration circular frequency $\omega$ is set as $2.6 \times 10^{6}\ rad/s$, which is consistent with our subsequent experiments. Elementary charge $q = 1.6 \times 10^{-19} C$, $\mu_n + \mu_p = 500\ cm^2/\mathrm{V}/s$, $\varepsilon = \varepsilon_r \varepsilon_0 = 8.854 \times 10^{-13} F/cm$, $f = 1$, absorption coefficient $\alpha = 0.001\ cm^{-1}$, quantum yield $\beta = 0.1$. For steady-state illumination, photon flux $\Phi$ is from $10^{3}\ cm^{-2}/s$ to $10^{20}\ cm^{-2}/s$. The direct recombination coefficients[10] $c_D$ is assumed to be $10^{-9} cm^3/s$. For trap-assisted recombination[8], $c_c = e_v$ is assumed to be $10^{-8}\ cm^3/s$. According to subsequent DFT calculations, the trap density $N_r$ is assumed to be $10^{7} cm^{-3}$, so $\tau_T = 20s$. For Auger recombination[7], $c_A$ is assumed to be $10^{-27} cm^6/s$. For sublinear recombination, $\gamma$ is set as 1/2, $c_{0.5}$ is assumed to be $10^{9} cm^{-3/2}/s$.

## Supplementary Note 7. Extract damping and resonance frequency from the measured amplitude-frequency curves

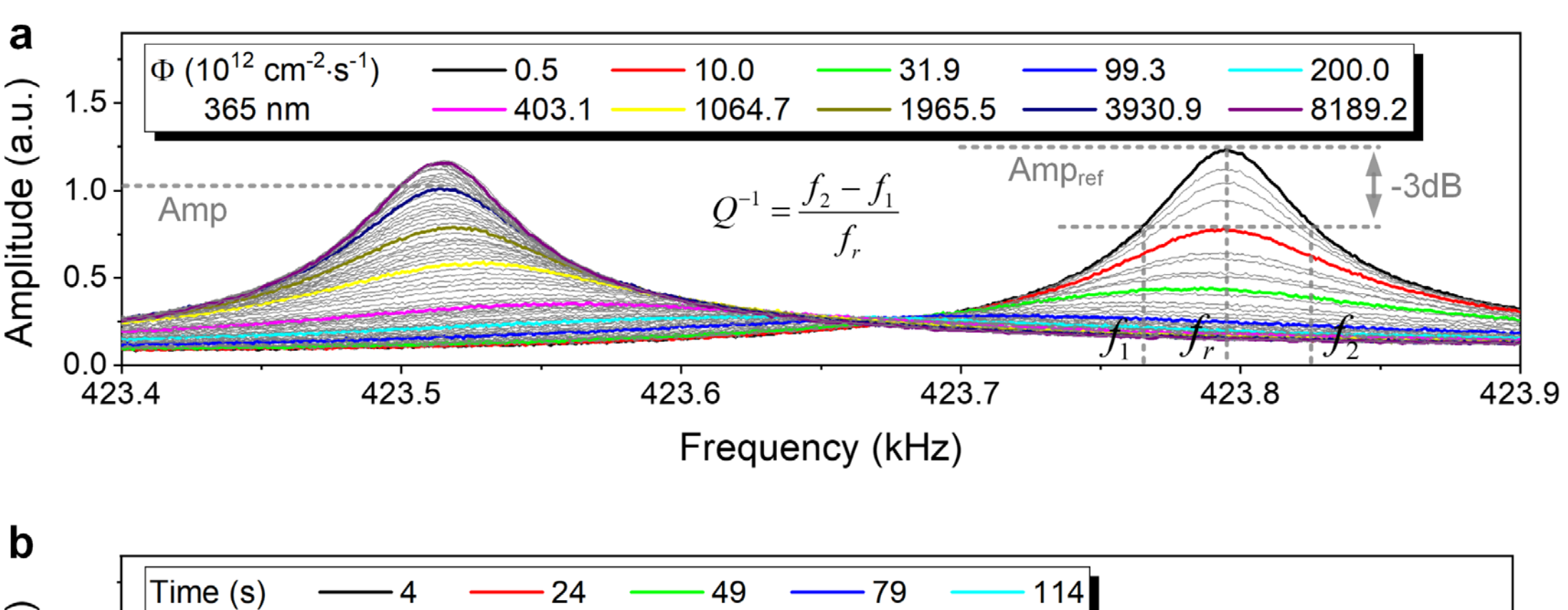

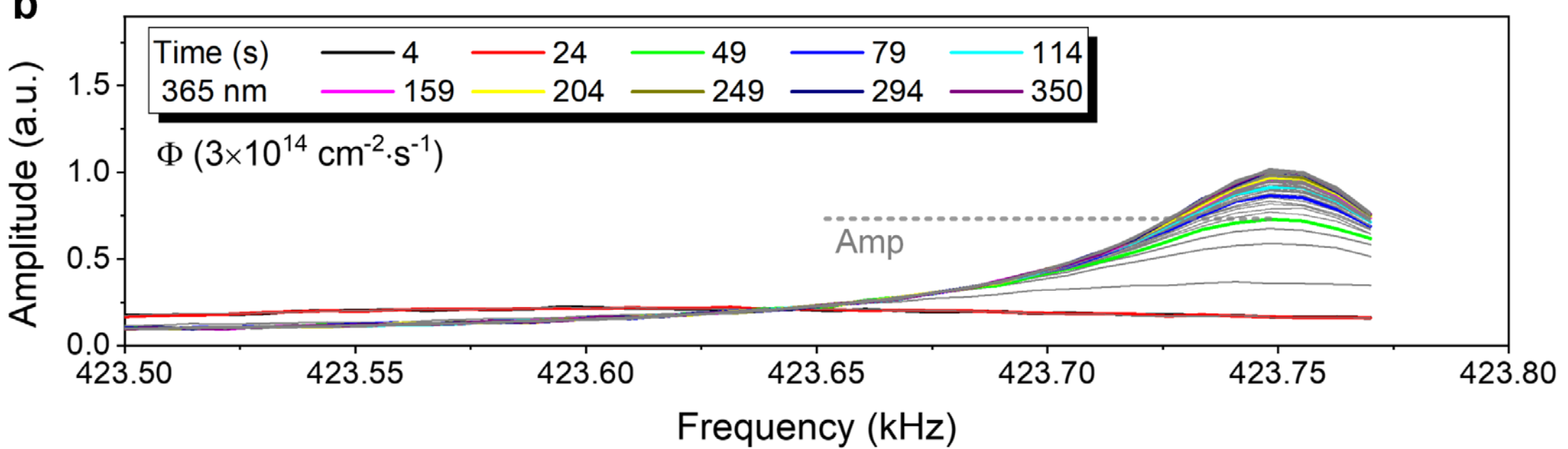

Supplementary Figure 7.1 Extract damping and resonance frequency from the measured amplitude-frequency curves under steady-state and transient illuminations. **a** Measured amplitude-frequency curves for ZnS (i) under steady-state 365 nm monochromatic light at various photon flux $\Phi$. **b** Time-dependent amplitude-frequency curves for ZnS (i) upon switching off 365 nm monochromatic light (with initial photon flux $\Phi = 3 \times 10^{14}\ cm^{-2} \cdot s^{-1}$).

Figure 7.1**a** shows the measured amplitude-frequency curves for ZnS (i) under steady-state 365 nm monochromatic light, with photon flux ranging from $10^{12}$ to $10^{17}\ cm^{-2} \cdot s^{-1}$. Under darkness, $\Phi$=0.5× $10^{12}\ cm^{-2} \cdot s^{-1}$, the $Q^{-1}$ can be calculated as[11]:

$$Q^{-1} = \frac{f_2 - f_1}{f_r} \tag{S22}$$

where $f_2$ and $f_1$ is derived from the half-bandwidth (-3 dB) of the peak, $f_r$ is determined from the peak position (Figure 7.1**a**).

For convenience, the $Q^{-1}$ under other illumination conditions can be calculated as[11]:

$$Q^{-1} = Q_{ref}^{-1} \frac{Amp_{ref}}{Amp} \tag{S23}$$

where $Q_{ref}^{-1}$ and $Amp_{ref}$ represent the damping and maximum value of the reference amplitude-frequency curves (i.e., the curve measured in darkness). While $Q^{-1}$ and $Amp$ denote the damping and maximum value of the amplitude-frequency curves measured under other illumination conditions. As shown in Figure 7.1**a**, with increasing photon flux, $Amp$ first decreases and then increases, indicating $Q^{-1}$ initially rises and subsequently falls, forming a peak in the $\Phi$-$Q^{-1}$ plot. Meanwhile, $f_r$ exhibits a transition from the unrelaxed to the relaxed state. Figure 7.1**b** shows that $Q^{-1}$ gradually recovers after the light is switched off.

## Supplementary Note 8. Fit the photon flux-damping and photon flux-resonance frequency change curve of ZnS using two models

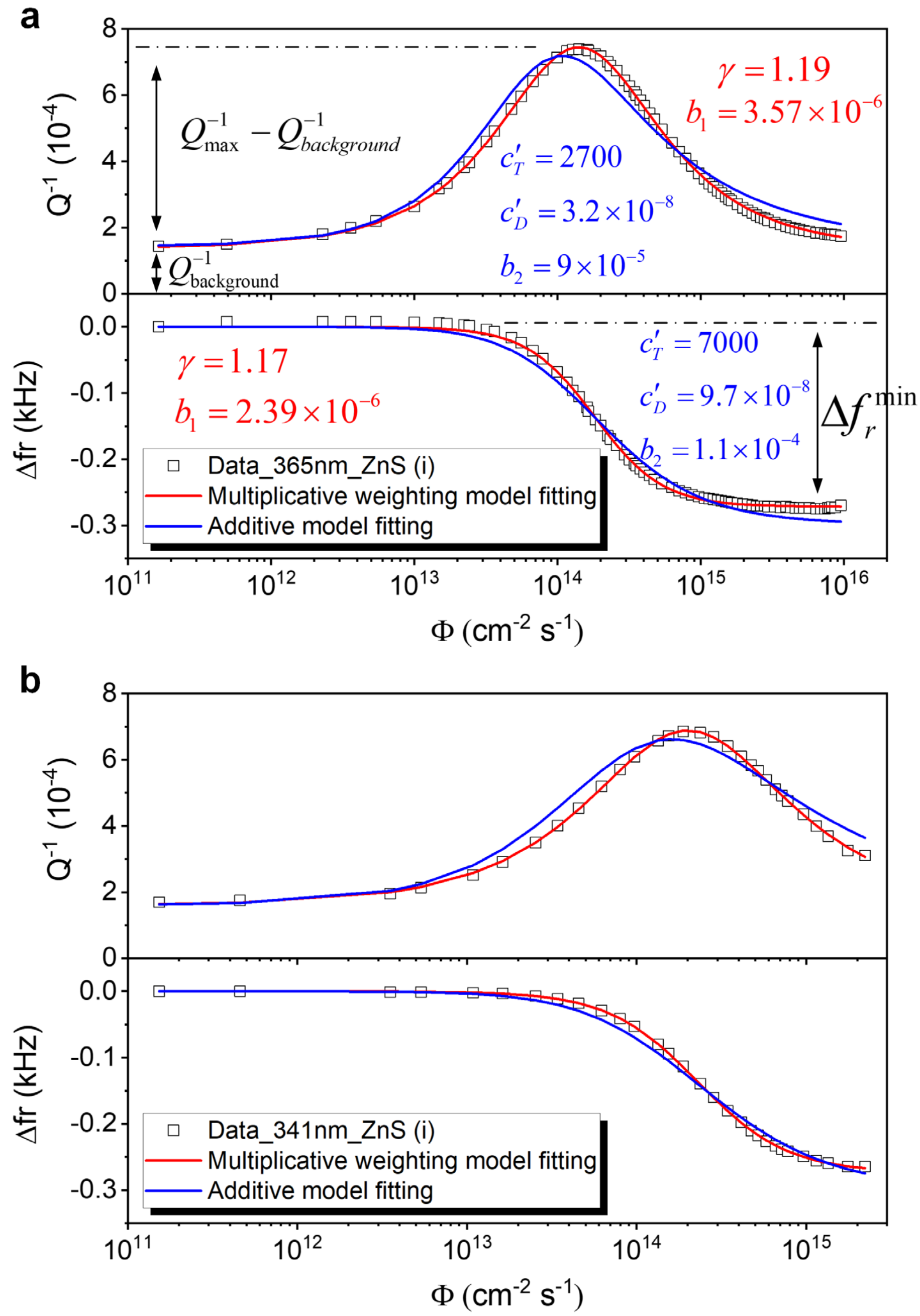

Supplementary Figure 8.1. Using the multiplicative weighting model and the conventional additive model to fit the steady-state $\Phi$-$Q^{-1}$ and $\Phi$-$\Delta f_r$ curves. **a** ZnS (i) under steady-state 365 nm monochromatic light. **b** ZnS (i) under steady-state 341 nm monochromatic light.

According to the amplitude-frequency curves measured under steady-state monochromatic light at various photon flux $\Phi$, we can obtain the $\Phi$-$Q^{-1}$ and $\Phi$-$\Delta f_r$ curves (Supplementary Note 7). Here we use the multiplicative weighting model and the conventional additive model to fit the $\Phi$-$Q^{-1}$ and $\Phi$-$\Delta f_r$ curves of ZnS, respectively.

For the multiplicative weighting model fitting, $\alpha\beta\Phi \propto (\Delta n)^{\gamma}$, so $\omega_c \propto \Delta n \propto \Phi^{1/\gamma}$:

$$\begin{cases} Q^{-1} = Q_{background}^{-1} + 2\times\left(Q_{\max}^{-1} - Q_{background}^{-1}\right)\dfrac{\omega_c/\omega}{1+\left(\omega_c/\omega\right)^2} \\ \Delta f_r = \Delta f_r^{\min}\dfrac{\left(\omega_c/\omega\right)^2}{1+\left(\omega_c/\omega\right)^2} \\ \omega_c = b_1\times\Phi^{1/\gamma} \end{cases} \tag{S24}$$

Where $Q_{background}^{-1}$ and $Q_{max}^{-1}$ represent the background and maximum mechanical damping, respectively. For $\Phi$-$Q^{-1}$ fitting, $Q_{background}^{-1}$, $Q_{max}^{-1}$, $b_1$, and $\gamma$ are the fitting parameters, while for $\Phi$-$\Delta f_r$ fitting, $\Delta f_r^{min}$, $b_1$, and $\gamma$ are the fitting parameters. The extracted $b_1$ and $\gamma$ values from the $\Phi$-$Q^{-1}$ and $\Phi$-$\Delta f_r$ curves are nearly identical (Figure 8.1**a**), validating both our theoretical framework for LIMAS and the applicability of the multiplicative weighting model to ZnS.

For the additive model fitting under steady-state 365 nm and 341 nm illumination, we consider the trap-assisted and direct recombination: $c_T\Delta n + c_D\Delta n^2 = \alpha\beta\Phi$, i.e., $c_T'\Delta n + c_D'\Delta n^2 = \Phi$

$$\begin{cases} Q^{-1} = Q_{background}^{-1} + 2\times\left(Q_{\max}^{-1} - Q_{background}^{-1}\right)\dfrac{\omega_c/\omega}{1+\left(\omega_c/\omega\right)^2} \\ \Delta f_r = \Delta f_r^{\min}\dfrac{\left(\omega_c/\omega\right)^2}{1+\left(\omega_c/\omega\right)^2} \\ \omega_c = b_2\times\dfrac{-c_T' + \sqrt{\left(c_T'\right)^2 + 4c_D'\Phi}}{2c_D'} \end{cases} \tag{S25}$$

Where $c_T$, $c_D$ are trap-assisted and direct recombination coefficients, respectively. For $\Phi$-$Q^{-1}$ curve fitting, $Q_{background}^{-1}$, $Q_{max}^{-1}$, $c_T'$, $c_D'$, $b_2$ are fitting parameters, while for $\Phi$-$\Delta f_r$ curve fitting, $\Delta f_r^{min}$, $c_T'$, $c_D'$, $b_2$ are fitting parameters.

Overall, the multiplicative weighting model provides an excellent fit to the experimental data, as shown by the red curves in Figure 8.1. In contrast, the additive model—despite including an additional fitting parameter—fails to accurately reproduce the data (blue curves in Figure 8.1), due to its inability to account for the coupling between recombination pathways.

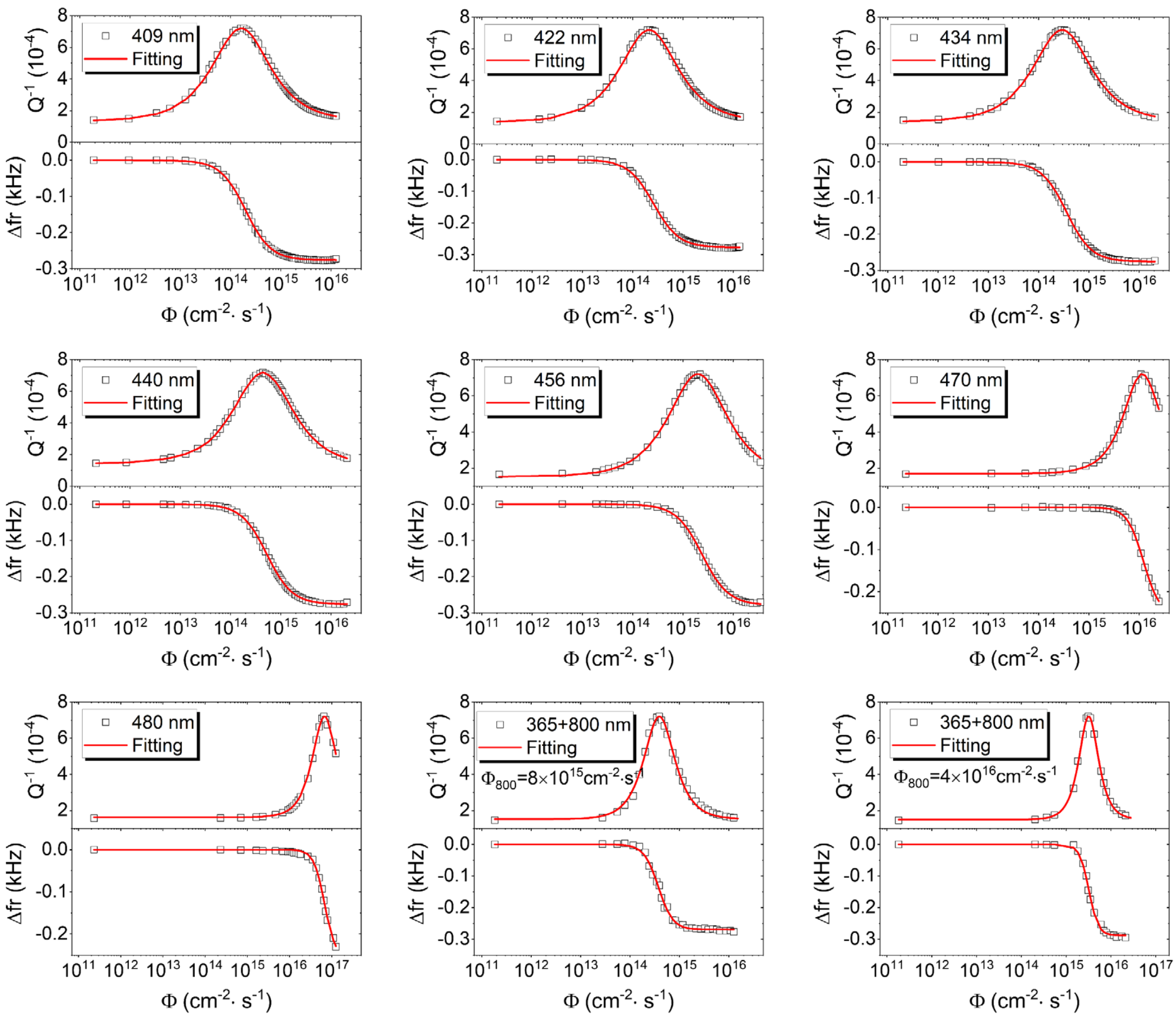

Supplementary Figure 8.2. Using the multiplicative weighting model to fit the steady-state $\Phi$-$Q^{-1}$ and $\Phi$-$\Delta f_r$ curves of ZnS (i) under different illumination conditions.

Figure 8.2 demonstrates the multiplicative model's robustness by providing consistently accurate fits to ZnS experimental data collected under various illumination conditions.

During the fitting, the corresponding photogenerated carrier concentration and resistivity of ZnS can be calculated by assuming reasonable parameters. Here we assume the carrier mobilities $\mu_n = \mu_p = 250\ cm^2/V/s$, the dielectric coefficient $\varepsilon = \varepsilon_r\varepsilon_0 = 8.854 \times 10^{-13} F/cm$, $f$=1, and the calculated results are shown in Figure 8.3. The photogenerated carrier density is in the range of $10^7$- $10^{13}$ cm$^{-3}$, and the corresponding resistivity is in the range of $10^8$-$10^3$ Ω·cm. These values are fully consistent with a high-resistivity ZnS crystal [12].

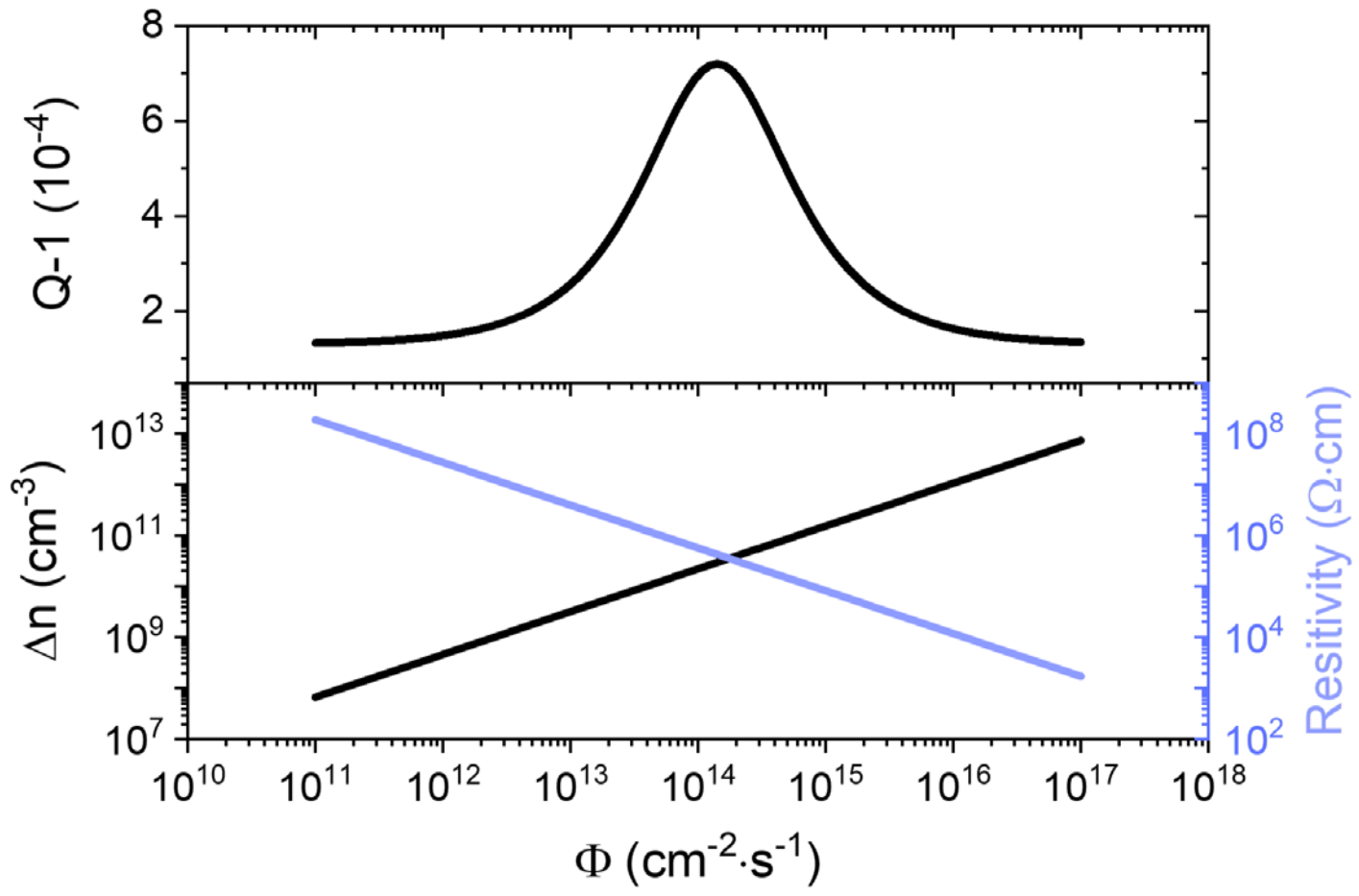

Supplementary Figure 8.3 Photogenerated carrier concentration extracted from a measured steady-state $\Phi$-$Q^{-1}$ curve and the corresponding resistivity

## Supplementary Note 9. Photon flux-damping and photon flux-resonance frequency change curves for GaN:C and GaAs:Cr

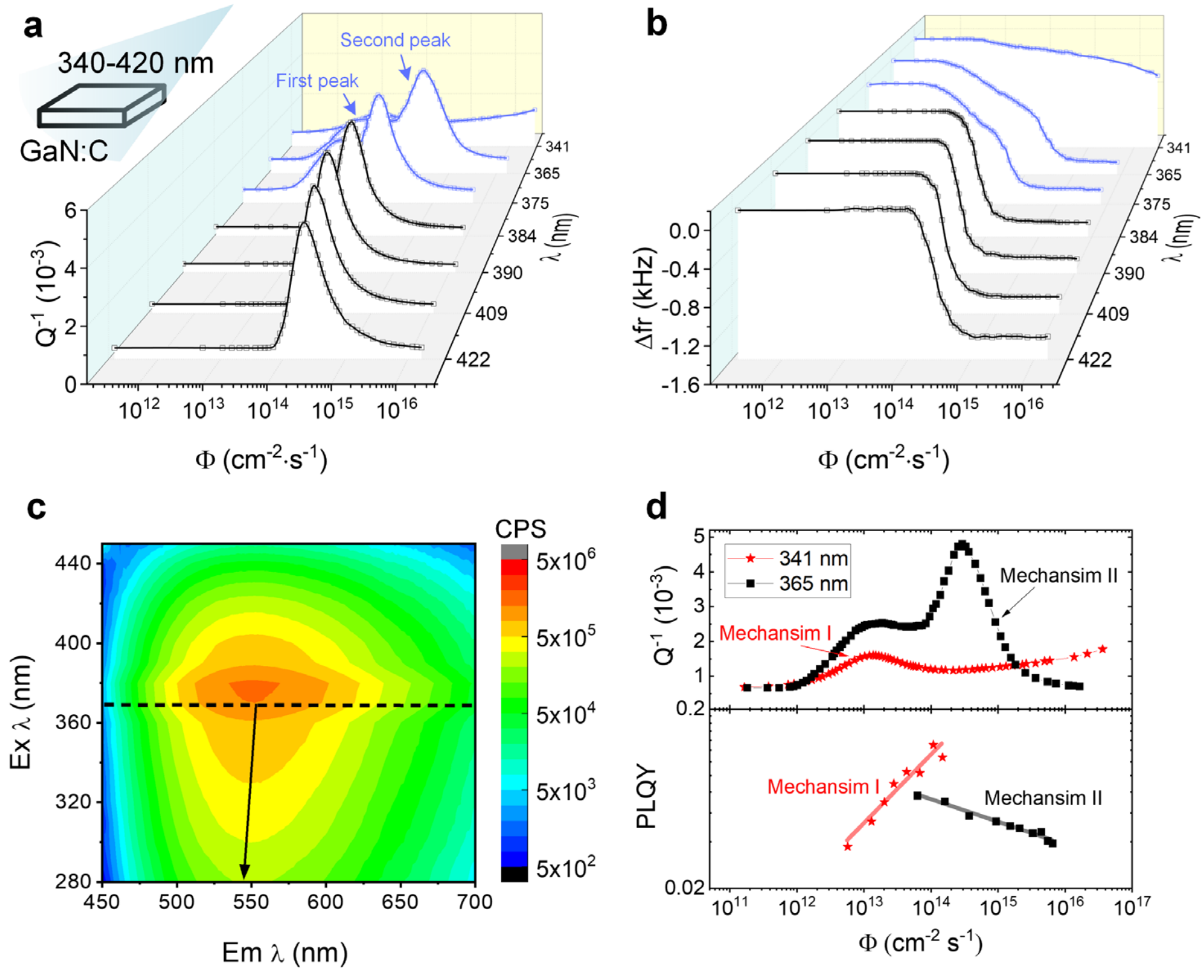

Supplementary Figure 9.1. Steady-state photon flux-damping (**a**), photon flux-resonance

frequency change curves (**b**), photoluminescence mapping (**c**), and PLQY (**d**) for carbon doped GaN

To further validate LIMAS and address the generality of our conclusions, we also performed tests on the III–V semiconductors. First, we measured the $\Phi$-$Q^{-1}$ and $\Phi$-$\Delta f_r$ curves of carbon doped GaN (GaN:C) under monochromatic light with the wavelengths from 340 nm to 420 nm (Figure 9.1 **a** and **b**). The resonance frequency of the sample used in these tests is about 304kHz. An interesting feature is that when the LED photon energy exceeds the sample bandgap (~375 nm, 3.3eV), two distinct peaks appear in the $\Phi$-$Q^{-1}$ curves (blue curves in Figure 9.1 **a** and black curve in Figure 9.1 **d**), accompanied by a shift in the photoluminescence peak (Figure 9.1 **c**). These two peaks exhibit different widths (Figure 9.1 **d**), and their corresponding PLQY trends (Figure 9.1 **d**) are also different (one increases with photon flux, whereas the other decreases), indicating the activation of two distinct recombination mechanisms in the different illumination conditions. The photon flux-dependent PLQY further demonstrates that GaN does not obey the multiplicative model. For comparison, we focus on fitting the $\Phi$-$Q^{-1}$ curves using the multiplicative ($p$→0) and additive model ($p$=1) in the wavelength range 384 nm - 420 nm, where only a single peak is observed. Since the $\Phi$-$Q^{-1}$ peak is very narrow, we must include a sublinear recombination pathway, which we model as $r_S = c_S \Delta n^{1/4}$ here. In addition, we consider a trap-assisted recombination pathway $r_T = c_T \Delta n$. So, the total rate can be written as: $r_{total} = r_S + r_T$ for additive model. The fitting procedures are almost the same as those described in Supplementary Note 8, and the results are summarized in Figure 9.2. For GaN:C, the $\Phi$-$Q^{-1}$ peaks are obviously asymmetric, and the additive model ($p$=1) better reproduces the experimental data than the multiplicative model ($p$→0), indicating a weak interaction between recombination pathways in GaN:C.

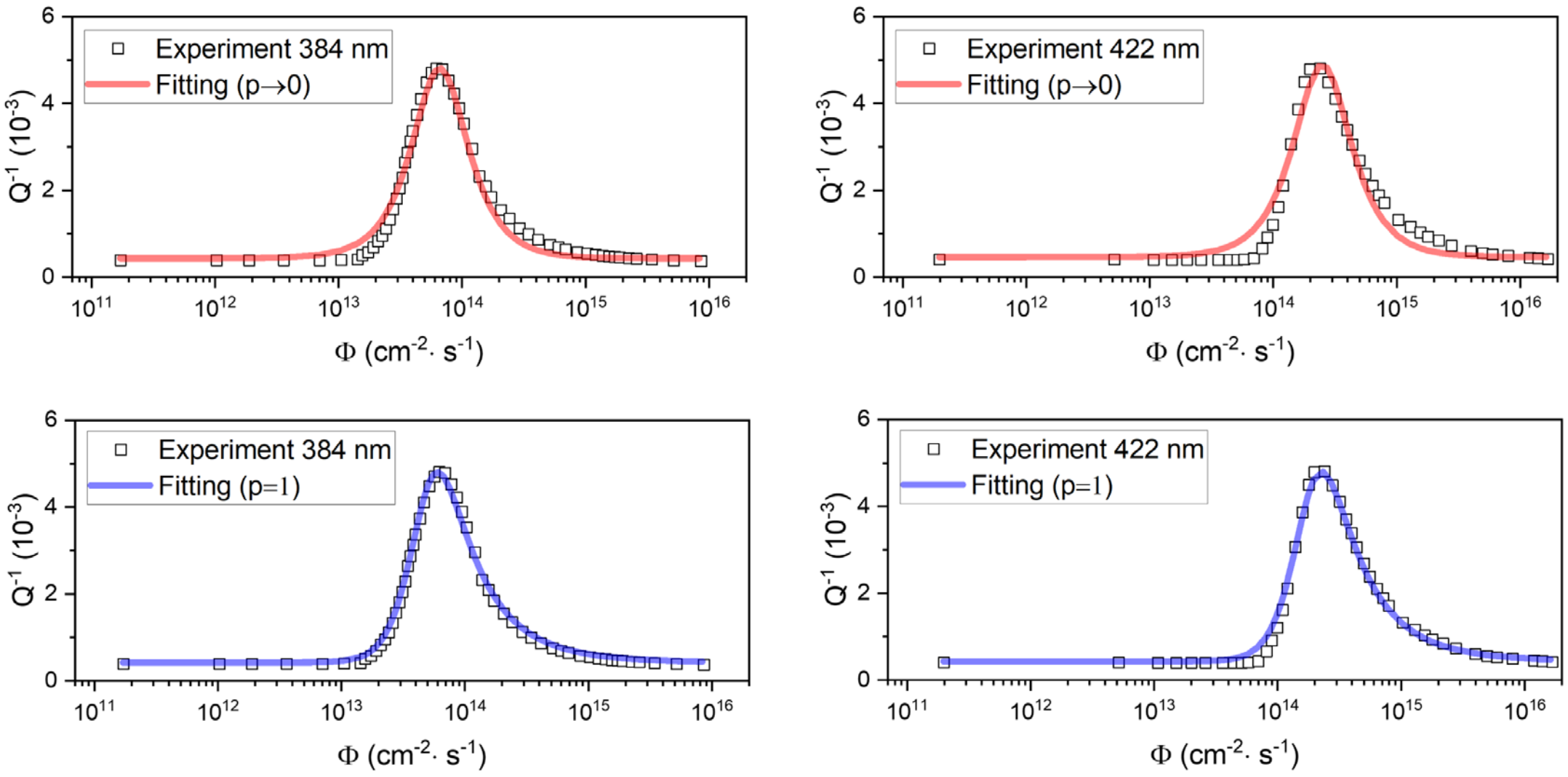

Supplementary Figure 9.2 Using the multiplicative weighting model ($p \to 0$) and the conventional additive model ($p$=1) to fit the steady-state $\Phi$-$Q^{-1}$ curves in GaN:C

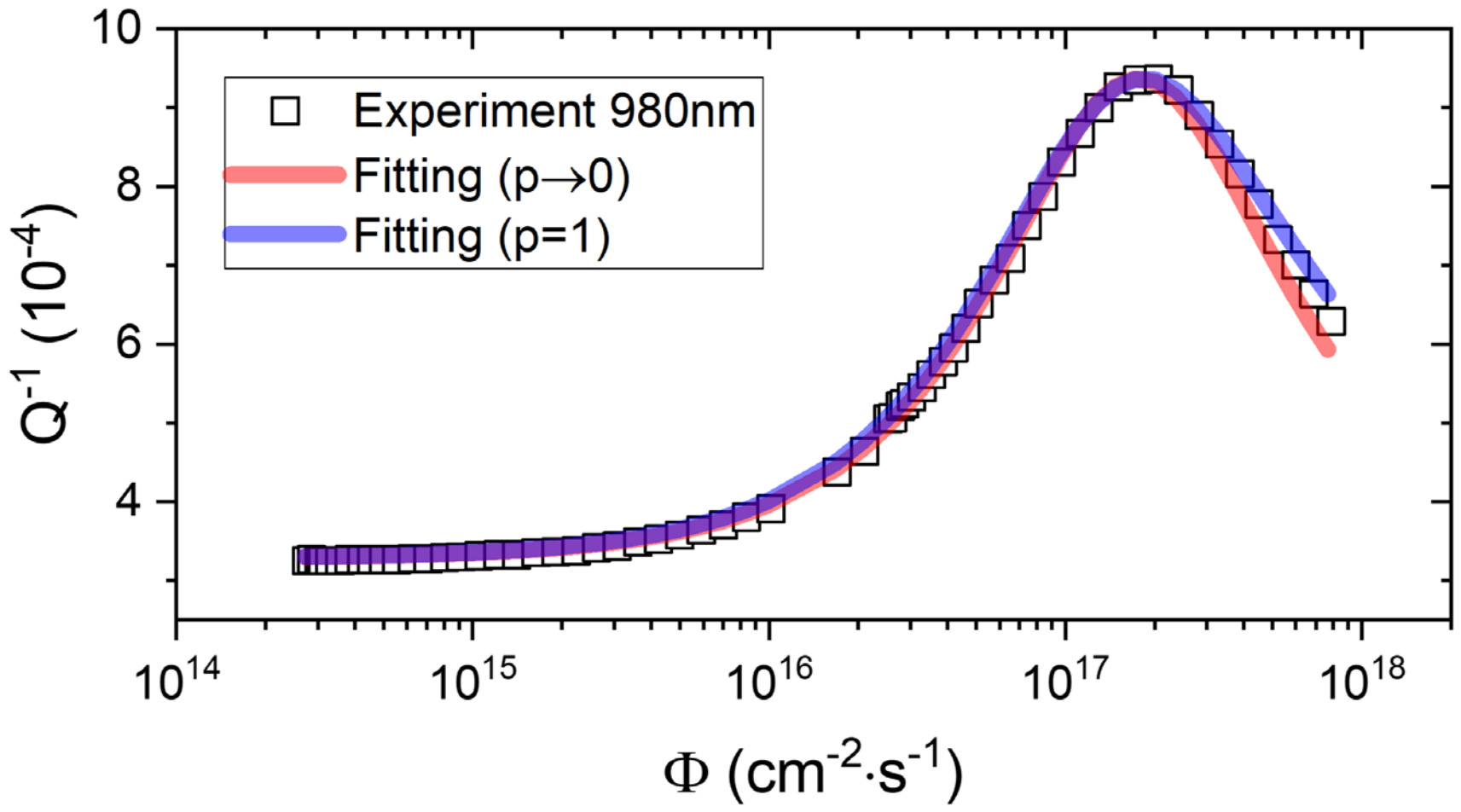

Supplementary Figure 9.3 Using the multiplicative weighting model ($p \to 0$) and the conventional additive model ($p$=1) to fit the steady-state $\Phi$-$Q^{-1}$ curves in GaAs:Cr

We also fitted the data for GaAs:Cr under 980 nm, with the results shown in Figure 9.3. The experimental data fall between the bounds of $p \to 0$ and $p$=1, indicating the values of $p \in (0,1]$ should be considered. These results for GaN and GaAs demonstrate LIMAS's generality.

## Supplementary Note 10. Photoluminescence quenching contour maps

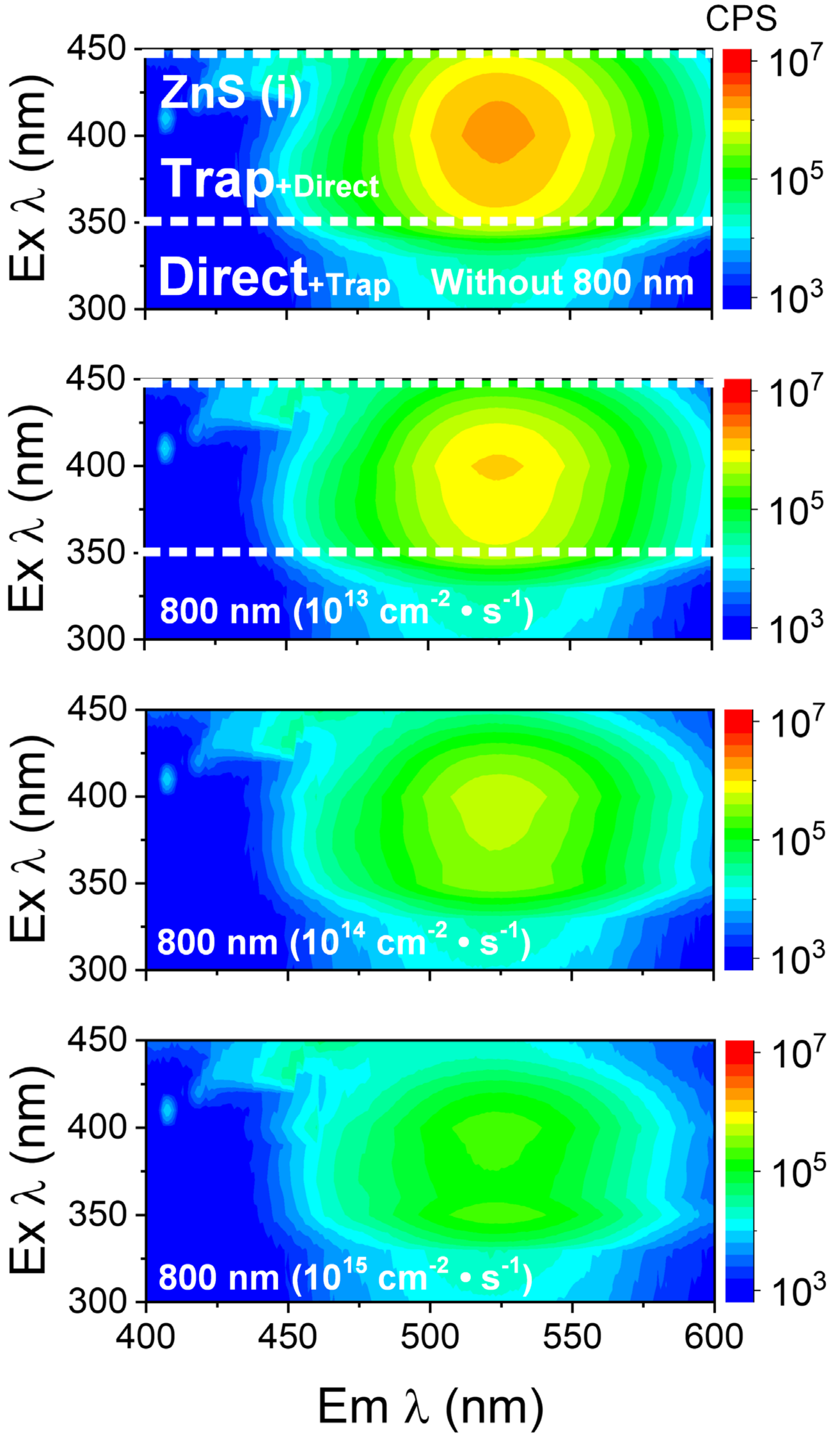

Supplementary Figure 10.1. Photoluminescence quenching contour maps of ZnS (i), measured using a spectrometer with excitation wavelengths ranging from 300 nm to 450 nm and detected emission wavelengths from 400 nm to 600 nm. An additional 800 nm LED with varying photon flux was used to induce quenching effects.

Figure 10.1 presents the photoluminescence contour map of ZnS (i), measured with excitation wavelengths ranging from 300 to 450 nm and emission collected from 400 to 600 nm. An additional 800 nm LED with variable photon flux was applied to induce quenching effects. A bandpass filter was used to eliminate any direct contribution from both the excitation source and the 800 nm illumination. The added 800 nm light reduces the intensity of the green emission without shifting its characteristic wavelength, indicating suppression of trap-assisted recombination. Based on this observation, 365 nm UV light was selected as the background excitation in Fig. 4, as it predominantly promotes trap-assisted recombination in ZnS.

## Supplementary Note 11. Resonance frequency change in ZnS under steady-state bichromatic illumination

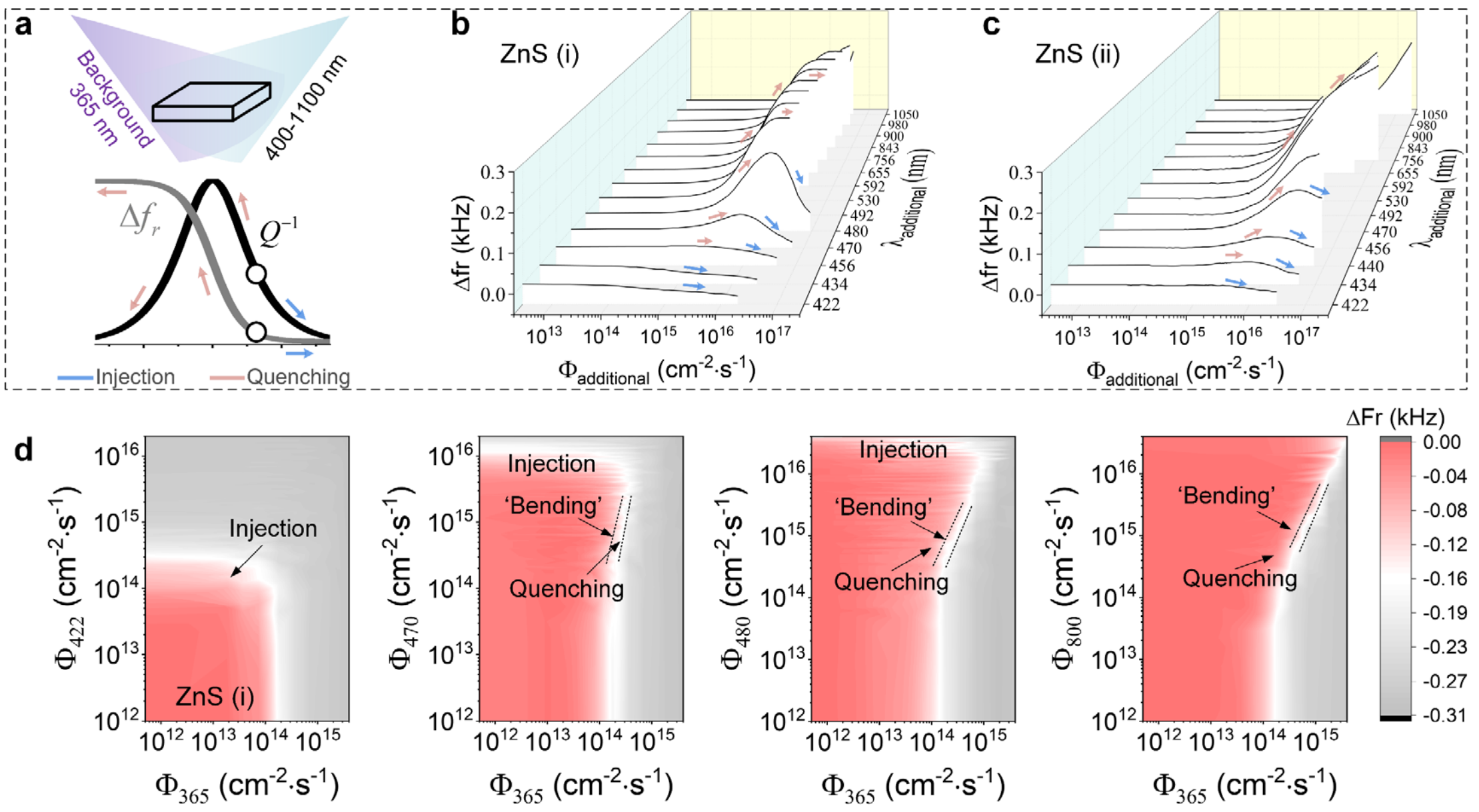

Supplementary Figure 11.1. Resonance frequency change in ZnS under steady-state bichromatic illumination. **a-c** Responses of $\Delta f_r$ in ZnS samples (i) and (ii) to constant-intensity 365 nm UV light, combined with an additional light source (400 nm-1100 nm) with varying photon flux ($10^{12}\ cm^{-2}\cdot s^{-1}$-$10^{17}\ cm^{-2}\cdot s^{-1}$). **d** Contour maps of $\Delta f_r$ in ZnS (i) under different bichromatic illumination conditions.

Figure 11.1 illustrates the resonance frequency change ($\Delta f_r$) in ZnS under steady-state bichromatic illumination. For ZnS (i), 365 nm UV light combined with additional light below 440 nm induces only injection (Figure 11.1**b**), leading to a decrease in $f_r$. When the additional light falls between 440 nm and 500 nm, injection and quenching compete, producing peaks in $\Delta f_r$ curves (Figure 11.1**b**). Beyond 520 nm, quenching dominates, causing an increase in $\Delta f_r$ (Figure 11.1**b**). ZnS (ii) follows similar trends, but quenching occurs at a shorter wavelength (~430 nm, Figure 11.1**c**). Figure 11.1**d** shows the contour maps of $\Delta f_r$ in ZnS (i) under different bichromatic illumination conditions. The 'bending' observed in the $\Delta f_r$ contour maps correspond to the quenching effect.

## Supplementary Note 12. Calculate time-dependent damping during the transient process

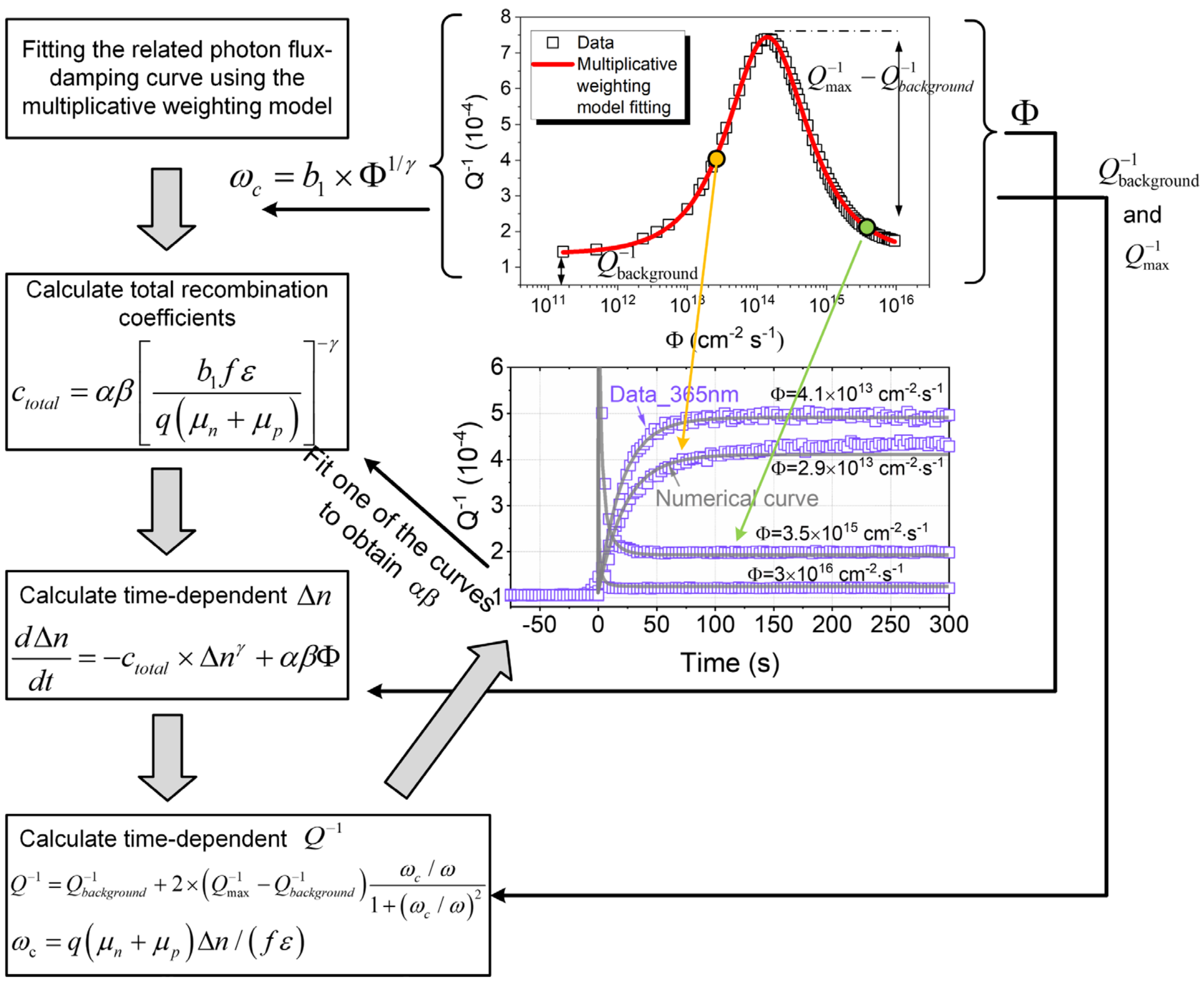

Supplementary Figure 12.1. Flowchart for calculating damping during the transient switch-on

process

Figure 12.1 presents a flowchart outlining the numerical procedure for calculating damping during the transient switch-on process. First, according to Supplementary Note 8:

$$\omega_c = \frac{q\left(\mu_n + \mu_p\right)}{f\varepsilon}\Delta n = b_1 \Phi^{1/\gamma} \tag{S26}$$

where $b_1$, $\gamma$, and $\Phi$ can be extracted from the measured steady-state $\Phi$-$Q^{-1}$ curve (Figure 12.1).

$$c_{total} \cdot \Delta n^{\gamma} = \alpha\beta\Phi \tag{S27}$$

and

$$c_{total} = \alpha\beta\left[\frac{b_1 f \varepsilon}{q\left(\mu_n + \mu_p\right)}\right]^{-\gamma} \tag{S28}$$

the parameters $\mu_n + \mu_p$, and $\varepsilon$ are taken from the literature (see Supplementary Note 6), and $f$ is set to 1. Once $\alpha\beta$ is specified, the time dependent $\Delta n$ can be calculated by solving the rate equation:

$$\frac{dn}{dt} = -c_{total} \cdot \Delta n^{\gamma} + \alpha\beta\Phi \tag{S29}$$

Then the time dependent $Q^{-1}$ can be obtained:

$$Q^{-1} = Q^{-1}_{background} + 2 \times \left(Q^{-1}_{\max} - Q^{-1}_{background}\right)\frac{\omega_c / \omega}{1 + \left(\omega_c / \omega\right)^2} \tag{S30}$$

By fitting the calculated $Q^{-1}$ to one of the experimentally measured curves, an appropriate value of $\alpha\beta$ can be identified. Once $\alpha\beta$ is fixed, varying the photon flux $\Phi$ allows determination of the time-dependent damping curves under different illumination conditions. To minimize the influence of trap occupancy dynamics (trap-assisted recombination, Supplementary Note 5), results under large photon flux (>$10^{13}$ $cm^{-2} \cdot s^{-1}$) should be used. The procedure for the switch-off case is the same and is not repeated here.

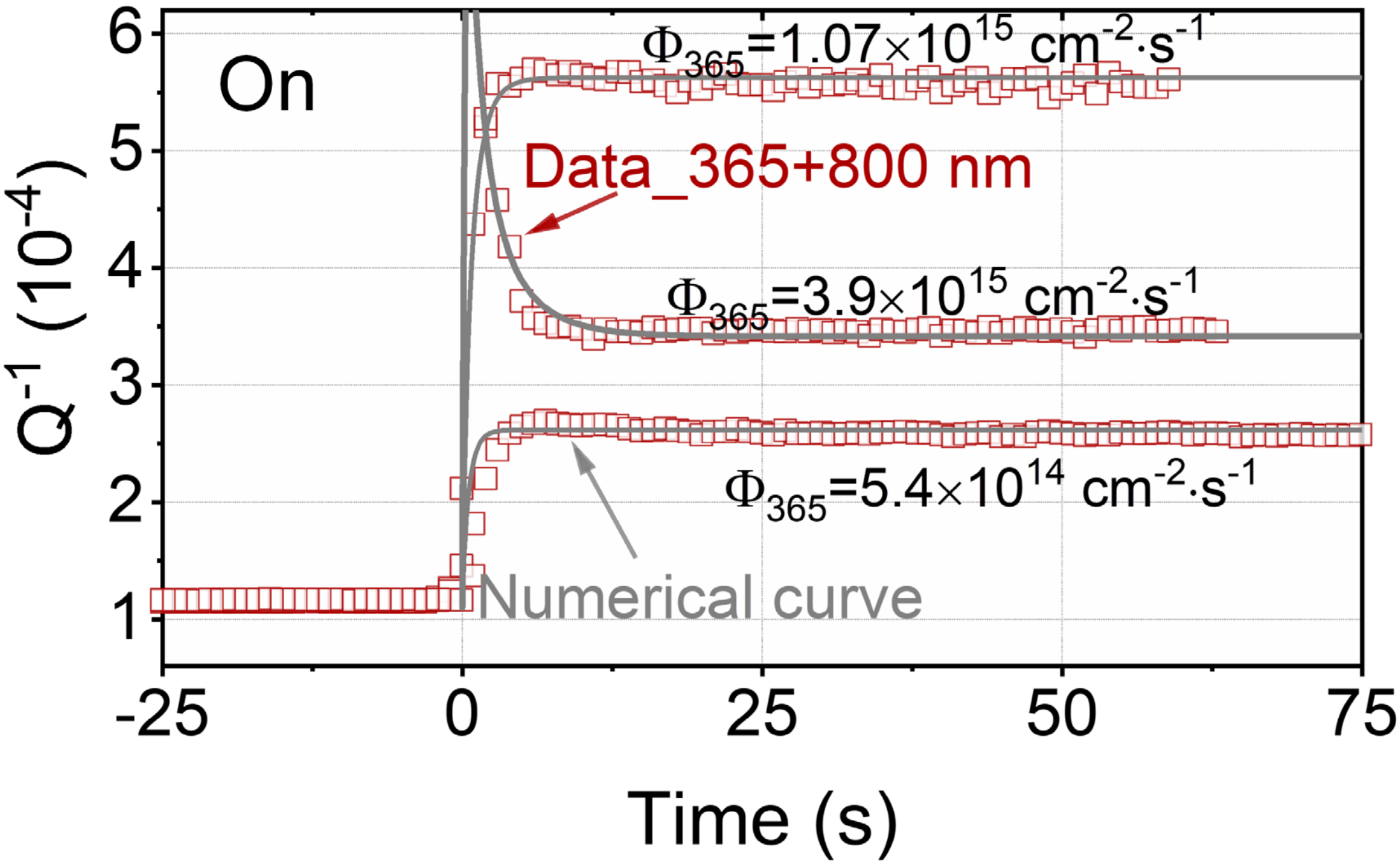

Supplementary Figure 12.2. Measured and numerically calculated (assuming constant $w_i$) time dependence of $Q^{-1}$ upon switching on of bichromatic light (365 nm + 800 nm, where the 800 nm light maintains open status and only 365 nm light at various photon flux is switched. $\Phi_{800nm} = 10^{15}\ cm^{-2} \cdot s^{-1}$).

Figure 12.2 shows a strong agreement between measured and numerically calculated (assuming constant $w_i$) time dependence of $Q^{-1}$ upon switching on of bichromatic light (365 nm + 800 nm, where the 800 nm light maintains open status and only 365 nm light at various photon flux is switched. $\Phi_{800nm} = 10^{15}\ cm^{-2} \cdot s^{-1}$), supporting our multiplicative weighting model, reflecting a constant-$\theta$ trajectory in Fig. 6**a**.

## Supplementary Note 13. Trap-assisted and sublinear recombination related defects

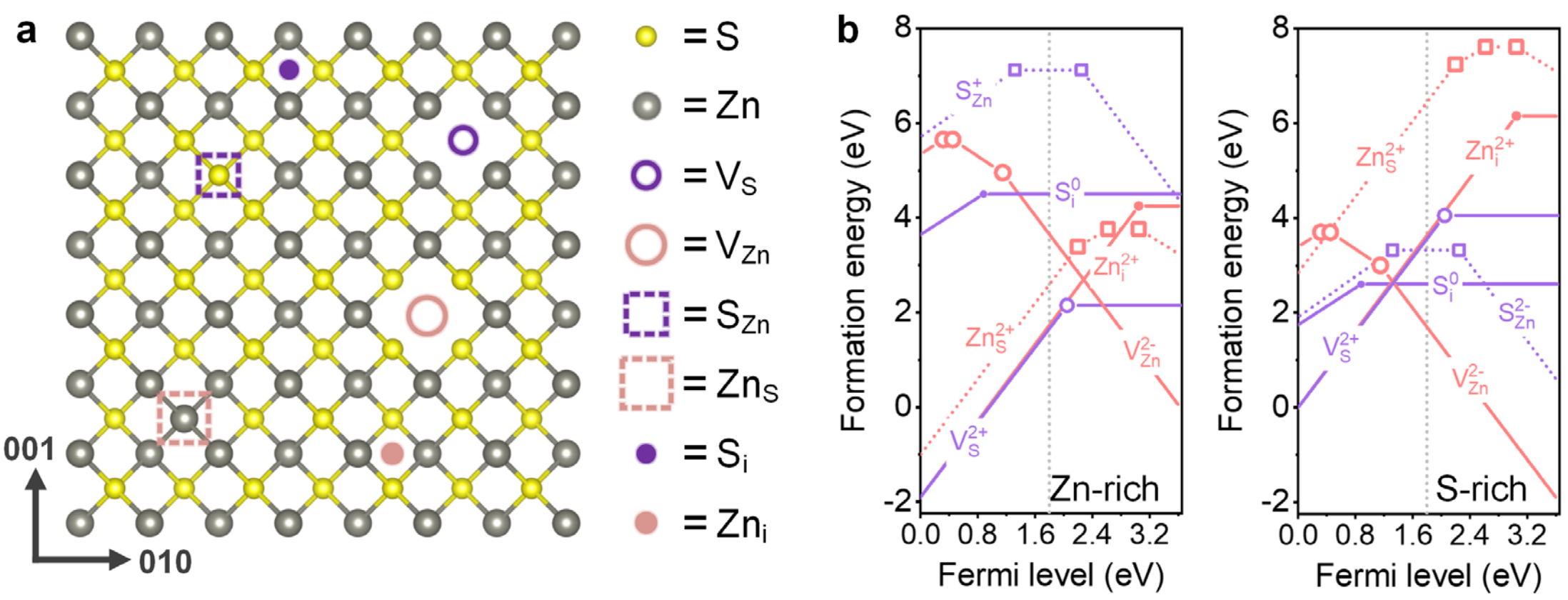

Supplementary Figure 13.1 Intrinsic point defects in ZnS associated with trap-assisted and sublinear recombination. **a** Intrinsic point defects in ZnS: vacancies ($V_S$, $V_{Zn}$), interstitials ($S_i$, $Zn_i$), and antisites ($S_{Zn}$, $Zn_S$). **b** Defect formation energy as a function of Fermi level relative to the valence band maximum.

ZnS contains various intrinsic point defects (Figure 13.1**a**), including vacancies ($V_S$, $V_{Zn}$), interstitials ($S_i$, $Zn_i$), and antisites ($S_{Zn}$, $Zn_S$). These defects act as donors or acceptors, with formation energies dependent on the growth environment[13-15]. Fig. 3 and 4 show that trap-assisted recombination in ZnS is linked to a donor defect more prevalent under Zn-rich conditions. This is evidenced by the stronger green photoluminescence in ZnS (i), which was annealed in vacuum, compared to ZnS (ii), which was annealed in a sulfur atmosphere. Based on the calculated formation energies in Figure 13.1**b**, zinc antisite ($Zn_S$) is identified as the trap-assisted recombination center. With charge transition points (2+/+ or +/0) near 2.4 eV, it effectively captures conduction band electrons, which then recombine with valence band holes, emitting green light (~2.4 eV). For sublinear recombination, a defect with a higher concentration than $Zn_S$ must serve as the initial electron reservoir at thermal equilibrium. Figure 13.1**b** identifies sulfur vacancy ($V_S$) as the most likely candidate. Its lower formation energy ensures a higher concentration[16], and its transition point (2+/0) near the center of the bandgap leads to full electron occupation at thermal equilibrium. Under low-energy monochromatic light (0.66 $Eg$-0.8 $Eg$) or with additional high-wavelength light (<0.8 $Eg$), electrons from $V_S$ are excited and transferred to $Zn_S$, progressively saturating both defect levels, leading to sublinear recombination, altering the emission properties.

The thermal equilibrium concentration of point defects can be calculated by:

$$N_r = N_{site} e^{\frac{-E_{Fr}}{k_0 T}} \tag{S31}$$

where $N_{site}$ is the number of lattice sites, approximately $10^{22}\ cm^{-3}$ for ZnS. $E_{Fr}$ is the formation energy. For zinc antisite in ZnS (i), assume the Fermi level is 1.6 eV, $E_{Fr}$ is about 0.9 eV, and the corresponding defect concentration $N_r$ is about $10^7\ cm^{-3}$.

All density functional theory (DFT) calculations are performed using the Vienna Ab initio Simulation Package (VASP) with the HSE hybrid functional. The screening length and the Hartree-Fock mixing parameter are set to 10Å and 0.32, respectively. The valence electron configuration considered are Zn: $3d^{10}4s^2$ and S: $3s^2 3p^4$, while inner electrons are treated as core states within the projector augmented wave (PAW) method. A plane-wave energy cutoff of 420 eV is used. A 64-atom supercell, consisting of 32 Zn and 32 S atoms, is employed to model intrinsic defects. Brillouin zone sampling is performed using a Gamma-centered 2×2×2 Monkhorst-Pack k-point mesh. Defect formation energies are determined using: $E_d^f = E_{defect} - E_{pristine} - \sum_i n_i u_i + qE_F$ . Where $E_{defect} - E_{pristine}$ represents the energy difference upon introducing a single defect in the structure. The remaining terms account for mass and charge compensation. $n_i$ is the added (positive) or removed (negative) number of atoms of species *i* with chemical potential $u_i$. $q$ is the defect charge state, and $E_F$ is the Fermi level.

## Supplementary Note 14. Power-mean model for carrier recombination in semiconductors

### Derivation

For a semiconductor with several coexisting recombination pathways (e.g., direct, trap-assisted, sublinear, Auger), each pathway can be characterized by a rate $r_i$ and a weight $w_i$ ($\sum w_i = 1$), and the total rate can be written using the power-mean model:

$$r_{total} \propto \left(\Sigma w_i r_i^p\right)^{1/p}, \quad \Sigma w_i = 1 \tag{S32}$$

For $p$=1, it can be reduced to the conventional additive model, $r_{total} \propto \sum w_i r_i$. For $p\to 0$, it reduces to the multiplicative weighting model: $\lim_{p\to 0} \ln(r_{total}) = \lim_{p\to 0} \frac{ln(\sum w_i r_i^p)}{p} = \lim_{p\to 0} \frac{ln\left[\sum w_i e^{p ln(r_i)}\right]}{p} = \lim_{p\to 0} \frac{d\left\{ln\left[\sum w_i e^{p ln(r_i)}\right]\right\}}{dp} = \lim_{p\to 0} \frac{\sum w_i ln(r_i) e^{p ln(r_i)}}{\sum w_i e^{p ln(r_i)}} = \frac{\sum w_i ln(r_i)}{\sum w_i} = \sum w_i \, ln(r_i)$, so we have:

$$r_{total} \propto \prod r_i^{w_i} \tag{S33}$$

Therefore, we reasonably infer that, for the general case, $p\epsilon(0,1]$. $p$ represents the strength of coupling (interaction) between recombination pathways.

### Physical meaning

The carrier recombination dynamics in semiconductors can be described within the framework of time-dependent perturbation theory of quantum mechanics:

$$\begin{cases} \hat{H}(t) = \hat{H}_0 + \hat{H}'(t) \\ \left|\psi(t)\right\rangle = \sum_n c_n(t) e^{-iE_n t/\hbar} \left|\phi_n\right\rangle \end{cases} \tag{S34}$$

where $\hat{H}_0$ is the time-independent Hamiltonian and $\hat{H}'(t)$ is a time-dependent perturbation associated with the recombination processes. Assuming that the system is initially in the $i$-th eigenstate $|i\rangle$, the first-order transition amplitude to a final state $|f\rangle$ is:

$$c_f^{(1)}(t) = \frac{1}{i\hbar}\int_0^t H'_{fi}(\tau) e^{i\omega_{fi}\tau} d\tau, \quad H'_{fi}(\tau) = \left\langle f \middle| \hat{H}'(\tau) \middle| i \right\rangle, \quad \omega_{fi} = \frac{E_f - E_i}{\hbar} \tag{S35}$$

and the corresponding transition probability is $P_{i\to f}(t) = \left|c_f^{(1)}(t)\right|^2$. For multiple coexisting pathways with different final states, the total transition probability is:

$$P_{total}(t) = \left\|\sum_f c_f^{(1)}(t)|f\rangle\right\|^2 = \sum_{f,f'} c_f^{(1)*}(t)c_{f'}^{(1)}(t)\langle f|f'\rangle \tag{S36}$$

and the total recombination rate is defined as:

$$R_{total} = \lim_{t\to\infty}\frac{P_{total}(t)}{t} \tag{S37}$$

When the final states form an orthonormal set, $\langle f|f'\rangle=\delta_{ff'}$, i.e., the pathways are decoupled, we have

$$R_{total} \propto \sum_f \left|c_f^{(1)}(t)\right|^2 = \sum R_f \tag{S38}$$

which automatically recovers the conventional additive model. However, if the final states are not orthogonal, indicating an effective coupling between different pathways. This coupling can occur when multiple pathways share intermediate or final states. In this case, the total recombination rate includes an extra interference term:

$$R_{total} \propto \sum_f \left|c_f^{(1)}(t)\right|^2 + \sum_{f\neq f'} c_f^{(1)*}(t)c_{f'}^{(1)}(t)\langle f|f'\rangle \tag{S39}$$

and the simple additive model is no longer valid. Such coupled behavior can instead be described phenomenologically by a generalized power-mean model. We emphasize that the power-mean model is a coarse-grained description of pathway coupling rather than a one-to-one rewriting of Eq. S39.

This coupling can also be visualized by an analogy of draining water from a pool (Figure 14.1). Each recombination pathway is represented by an outlet with a discharge rate ($r_i$). Within the power-mean model, when $p$=1, the outlets are effectively independent: they are far apart, do not interfere with one another, and the total flow is essentially the simple sum of the individual rates. In contrast, as $p\to0$, the outlets are very close, their vortices strongly interact, and the total discharge rate is no longer given by a simple additive sum. Thus, $p \in (0,1]$ serves as a parameter that quantifies the degree of coupling between recombination pathways.

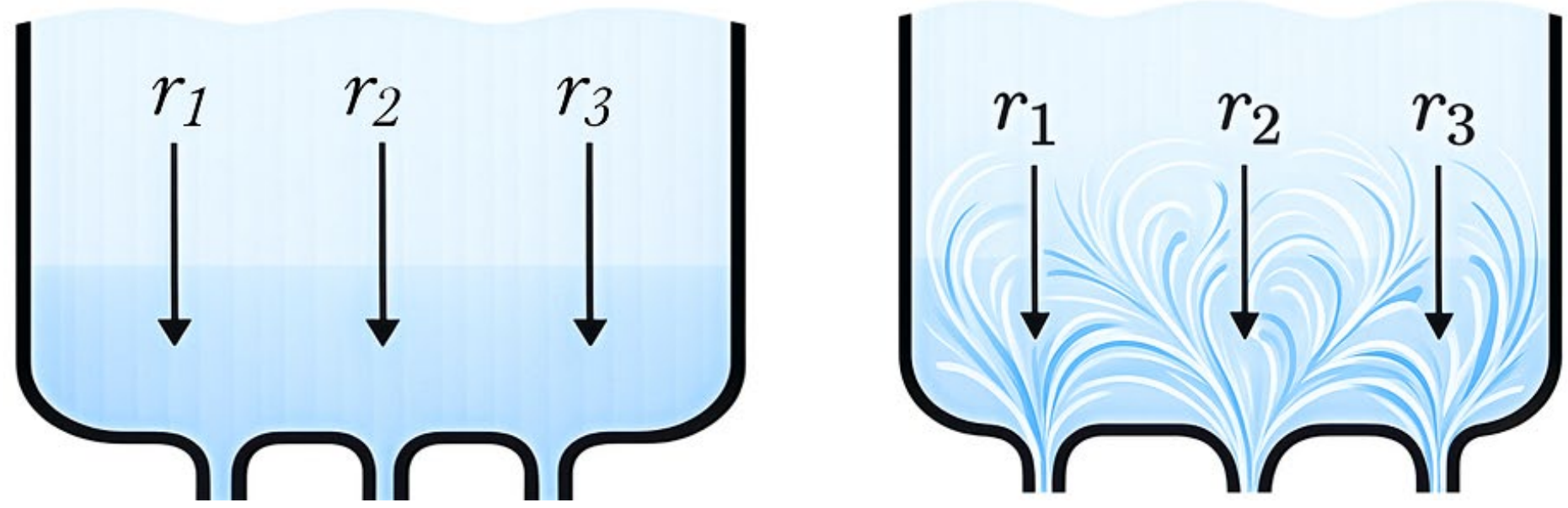

Supplementary Figure 14.1 Consider the recombination process as draining water from a pool, with each pathway representing an outlet with a discharge rate $r_i$. When $p = 1$, the interactions between outlets vanish. In contrast, as $p \to 0$, the pathways become strongly coupled. Thus $p \in (0,1]$ quantifies the degree of coupling.

The power-mean model, and more broadly, the concept of introducing a parameter $p$ to generate a family of models that interpolates between different limiting cases, is widely used in mathematics and physics, including:

$L^p$ norms:

For a vector $x$, the $L^p$ norm

$$\|x\|_p = \left(\sum |x_i|^p\right)^{1/p} \tag{S40}$$

interpolates between the maximum norm $\|x\|_\infty = max(x_i)$ as $p \to \infty$, and the sum norm $\|x\|_1 = \sum x_i$ at $p$=1. Thus the parameter $p$ controls how strongly the largest component dominates the average.

$p$-Laplacian operators:

$$\Delta_p u = \nabla \cdot \left( |\nabla u|^{p-2} \nabla u \right) \tag{S41}$$

generalizes the usual operator: for $p$=2 it reduces to the standard diffusion operator $\Delta$u, while other values of $p$ describe nonlinear diffusion where high gradient regions are weighted differently.

Fractional Schrödinger equation [17]:

$$i\hbar \frac{\partial \psi(x,t)}{\partial t} = \left[ D_\alpha \left(-\hbar^2 \Delta\right)^{\alpha/2} + V(x,t) \right] \psi(x,t), \quad 0<\alpha \le 2 \tag{S42}$$

The parameter α interpolates between conventional quantum dynamics (α=2) and more nonlocal Levy-flight-like behavior for α<2.

Table 14.1 lists some power-mean model in different physical systems.

**Table 14.1 Examples of power-mean models in different physical systems**

| Physics system | Power-mean models |
|---|---|
| Mean elastic modulus in nontextured polycrystal [18] | Voigt model (uniform strain, upper bound, p=1);<br>Reuss model (uniform strain, lower bound, p=-1);<br>Hill model (intermediate, -1<p<1) |
| Effective thermal conductivity in layered composites | In-plane (parallel heat paths, p=1);<br>Cross-plane (series heat path, p=-1); |
| Electrical resistor networks | Series resistors (p=1);<br>Parallel resistors (p=-1);<br>2D self-dual two-phase resistor (p→0) [19, 20] |
| **Carrier recombination dynamics** | **Conventional ABC model (p=1);**<br>**Multiplicative model (p→0)** |

Then we will discuss the physical meaning of the power-mean models in carrier recombination dynamics from the perspective of energy. The total recombination rate can be written as:

$$r_{total}\left(T,n;p\right)\propto\left[\Sigma w_i r_i\left(T,n\right)^p\right]^{1/p},\quad \Sigma w_i=1 \tag{S43}$$

where

$$r_i(T,n)=A_i e^{\frac{-E_i}{K_BT}} f_i\left(n\right) \tag{S44}$$

with $E_i$ the activation energy of the *i*-th pathway and $f_i(n)$ a factor that depends only on carrier density *n.* The effective activation energy extracted from the Arrhenius slope is [21]:

$$E_{eff}\left(T,n;p\right)=-K_B\left.\frac{\partial\ln r_{total}(T,n;p)}{\partial(1/T)}\right|_n=\frac{\sum w_i A_i^p e^{\frac{-pE_i}{K_BT}} f_i\left(n\right)^p E_i}{\sum w_j A_j^p e^{\frac{-pE_j}{K_BT}} f_j\left(n\right)^p} \tag{S45}$$

For $p$=1, $E_{eff}=\frac{\Sigma w_i r_i(T,n)E_i}{\Sigma w_j r_j(T,n)}$, which is simply a rate-weight average of the individual activation energies $E_i$ [21]. As the carrier density changes, $r_i(T,n)$ change strongly, and $E_{eff}$ is therefore expected to drift between the different $E_i$. For $p$→0, $E_{eff}=\frac{\Sigma w_i E_i}{\Sigma w_j}$, which is a constant structure average independent of $T$ and $n$. Mathematically, the two limits $p$=1 and $p$→0 resemble annealed (rate-weighted) and quenched (structural) averages in statistical physics, respectively. Table 14.2 summarizes the expressions for several commonly used quantities in optoelectronics using the power-mean model, as well as the related two limiting cases.

**Table 14.2 Physical quantities in optoelectronics within the power-mean framework**

| Physical quantity | $p\epsilon(0,1]$ | $p$=1 | $p$→0 |
|---|---|---|---|
| $r_{total}$ | $\propto \left(\sum w_i r_i^p\right)^{1/p}$ | $\propto \sum w_i r_i$ | $\propto \prod r_i^{w_i}$ |
| $E_{eff}$ | $\frac{\sum w_i r_i^p E_i}{\sum w_j r_j^p}$ | $\frac{\sum w_i r_i E_i}{\sum w_j r_j}$ | $\frac{\sum w_i E_i}{\sum w_j}$ |
| PLQY ($\eta_{PL}$) | $\frac{\sum w_i r_i^p \chi_i}{\sum w_j r_j^p}$ | $\frac{\sum w_i r_i \chi_i}{\sum w_j r_j}$ | $\frac{\sum w_i \chi_i}{\sum w_j}$ |

Here $\chi_i\epsilon[0,1]$ denotes the probability that a recombination event through the $i$-th pathway produces a photon. $\frac{w_i r_i^p}{\sum w_j r_j^p}$ is the normalized participation factor (i.e., the logarithmic sensitivity $\frac{\partial ln(r_{total})}{\partial ln(r_i)}$) of the $i$-th pathway, it measures how much a small relative change of $r_i$ (e.g. by 1%) contributes to the relative change of the total rate, and can reduce to the usual fractional contribution $\frac{w_i r_i}{\sum w_j r_j}$ in the additive limit (p=1). It should be noted that, for $p$→0, the PLQY becomes intensity-independent.

## Experiments to support the power-mean framework

Intensity-independent PLQY in ZnS

For ZnS, the mechanical damping measurements indicate that the recombination dynamics are well described by the multiplicative weighting model, a limit case of the power-mean framework ($p$→0). In this limit, the theory predicts that the PLQY should be independent of photon flux or carrier density (Table 14.2). To better validate this, we measured the PLQY of sample ZnS (i) (which has a higher photoluminescence than ZnS (ii), making it more suitable for PLQY tests) using an integrating sphere, LEDs (340 nm and 365 nm), a photodetector and two bandpass filters. For the absorption measurement, a UV bandpass filter (FGUV11, 275 nm–375 nm, Thorlabs) was used to block the green emission (520 nm) from the ZnS sample, whereas for the emission measurement a second bandpass filter (FGV9S, 485 nm–565 nm, Thorlabs) was used to block the UV light from the LEDs. The experimental setup and results are shown in Figure 14.2. As can be seen, the PLQY remains nearly constant over the entire range of the excitation photon flux, in agreement with the $p \to 0$ prediction. This result is in sharp contrast to the strongly intensity-dependent PLQY described by the conventional ABC-type models [22]. Combining the PLQY data here with the weighting factors $w_i$ obtained from the manuscript Fig. 3, and using equation in Table 14.2, we estimate that in ZnS (i), the luminous probability of the trap-assisted recombination pathway, $\chi_{trap}$, at 365 nm is approximately 5%.

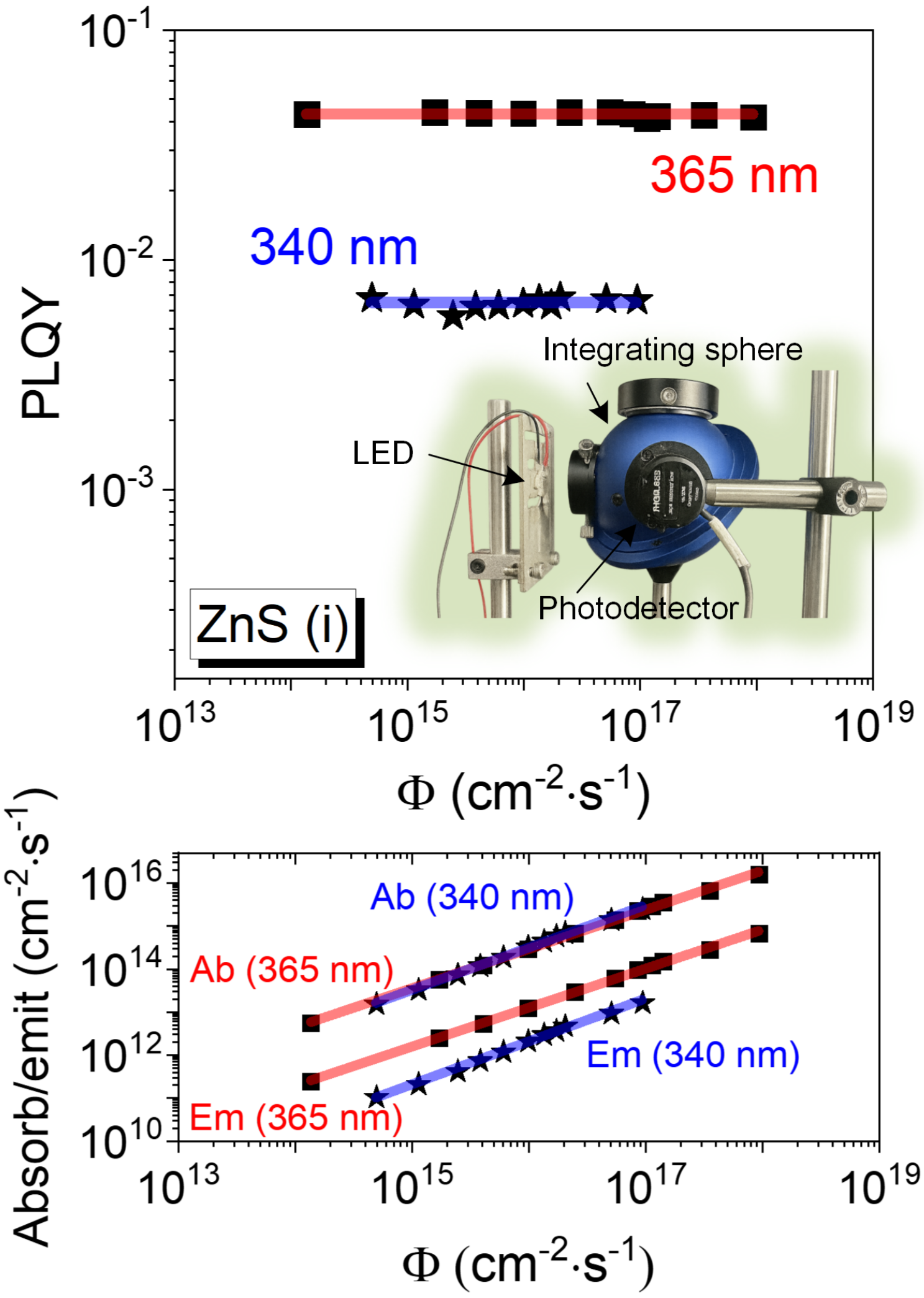

Supplementary Figure 14.2 Experimental setup and results of photon flux-dependent PLQY in ZnS (i) at monochromatic excitation 340 nm and 365 nm

Other experiments in the literature

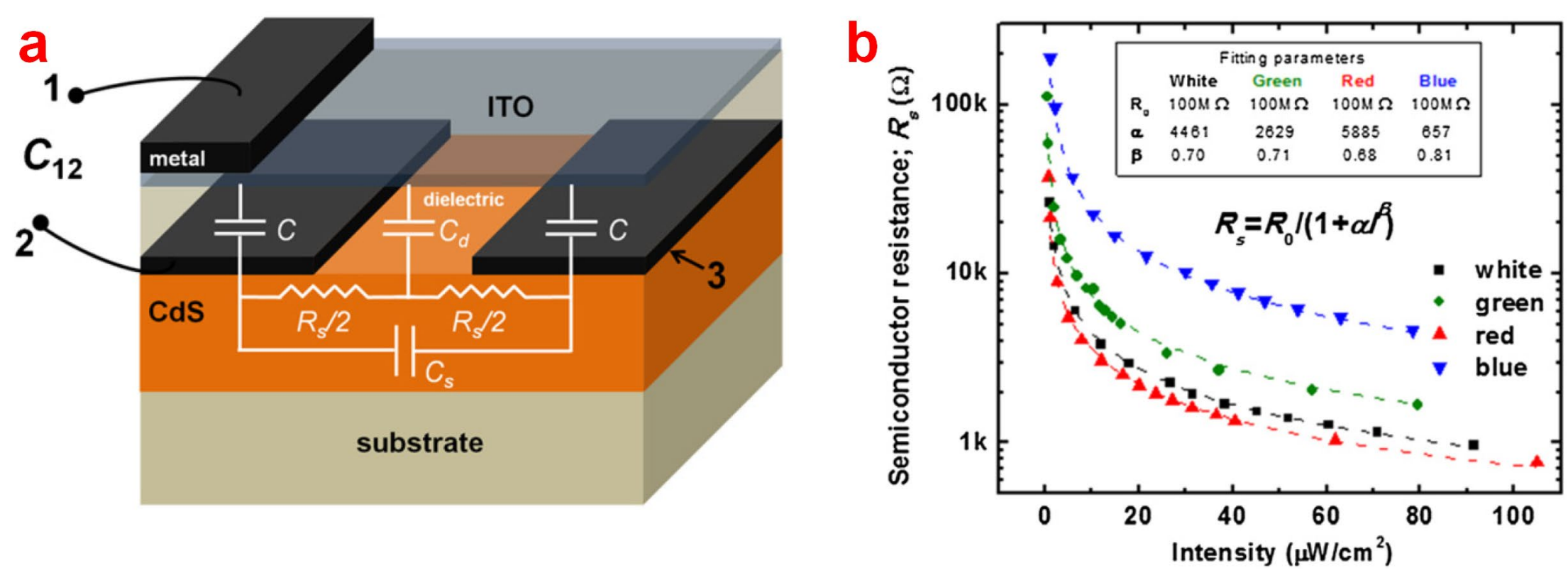

Supplementary Figure 14.3 Photo-resistance of a CdS-based light sensor [23]

In the work of Saxena et al.[23], the photo-resistance (Figure 14.3 **b**) of a CdS-based light sensor (Figure 14.3 **a**) was measured. They discovered that the change in conductivity (reciprocal of resistivity) follows a power law with the photon flux:

$$\left(R_s^{-1} - R_0^{-1}\right) \propto \Delta n \propto \Phi^{\beta} \tag{S46}$$

the exponent $\beta$ varies with illumination conditions. This behavior strongly supports our multiplicative model ($p \to 0$), indicating that such multiplicative behavior is not unique to ZnS. It further suggests that previous experiments may have overlooked certain information and may be fruitfully revisited within the power-mean framework.